\documentclass[aps,prc,showpacs,showkeys,superscriptaddress,nofootinbib,twocolumn,floatfix]{revtex4}

\usepackage{bm}
\usepackage{placeins}
\usepackage{graphicx}
\usepackage{amsfonts}
\usepackage{bbm}
\usepackage{multirow}
\usepackage{graphicx,color,amsmath,amssymb}

\usepackage{hyperref}
\hypersetup{
            colorlinks=true,
            linkcolor=blue,
            anchorcolor=blue,
            citecolor=blue}

\def\eg{{\em e.g.}}
\def\ie{{\em i.e.}}

\newcommand{\beq}{\begin{equation}}
\newcommand{\eeq}{\end{equation}}
\newcommand{\bea}{\begin{eqnarray}}
\newcommand{\eea}{\end{eqnarray}}

\newcommand{\gtsim}{\raisebox{-4pt}{$\,\stackrel{\textstyle >}{\sim}\,$}}

\begin{document}

\title{Spin-Dependent Interactions and Heavy-Quark Transport in the QGP}

\author{Zhanduo Tang} \author{Ralf Rapp}
\affiliation{Cyclotron Institute and Department of Physics and Astronomy,
Texas A\&M University, College Station, TX
  77843-3366, USA}

\date{\today}

\begin{abstract}
We extend a previously constructed $T$-matrix approach to the quark-gluon plasma (QGP) to include the effects of spin-dependent interactions between partons. Following earlier work within the relativistic quark model, the spin-dependent interactions figure as relativistic corrections to the Cornell potential. When applied to the vacuum spectroscopy of quarkonia, in particular their mass splittings in $S$- and $P$-wave states, the issue of the Lorentz structure of the confining potential arises. We confirm that a significant admixture of a vector interaction (to the previously assumed scalar interaction) improves the description of the experimental mass splittings.  
The temperature corrections to the in-medium potential are constrained by results from thermal lattice-QCD for the equation of state (EoS) and heavy-quark (HQ) free energy in a selfconsistent set-up for heavy- and light-parton spectral functions in the QGP. We then deploy the refined in-medium 
heavy-light $T$-matrix to compute the charm-quark transport coefficients 
in the QGP.
The vector component of the confining potential, through its relativistic corrections, enhances the friction coefficient for charm quarks in the QGP 
over previous calculations by tens of percent at low momenta and temperatures, and more at higher momenta. Our results are promising for 
improving the current phenomenology of open heavy-flavor observables at Relativistic Heavy Ion Collider (RHIC) and the Large Hadron Collider (LHC).
\end{abstract}

\maketitle


\section{Introduction} 
\label{sec_intro} 
The exploration of hadron properties in vacuum and the properties of the quark-gluon plasma (QGP) are usually regarded as rather independent areas in the study of Quantum Chromodynamics (QCD). However, in both areas the basic building block are soft parton interactions rooted in the non-perturbative sector of the theory, albeit in different environments. Of particular interest are heavy quarks: heavy-quarkonium spectroscopy in vacuum has provided deep insights into potential between a heavy (charm or bottom) quark ($Q=c,b$) and its antiquark ($\bar Q$). The Cornell potential and its refinements remain a phenomenologically successful tool in the description of the pertinent bound states, taking advantage of expansion in the inverse heavy-quark (HQ) mass, $1/M_Q$~\cite{QuarkoniumWorkingGroup:2004kpm}. The long-range (linear) part of the potential, which by now is also well established in lattice QCD (lQCD), is arguably one of the most direct manifestations of the confining force in QCD. In the context of high-temperature QCD and its study in ultrarelativistic heavy-ion collisions (URHICs), this led to the idea of utilizing quarkonium production as a probe of deconfinement, although the originally proposed suppression signature has evolved considerably over the last three decades~\cite{Rapp:2008tf,Kluberg:2009wc,Braun-Munzinger:2009dzl}. 
Specifically, transport approaches have been developed toward the more general objective of deducing the in-medium QCD force from quarkonium observables, 
by implementing it into the transport coefficients that govern both suppression 
and regeneration reactions in the evolving fireball of a heavy-ion collision, see, \eg, Ref.~\cite{Du:2019tjf}. This effort is critically aided by ample information from lQCD on the in-medium properties of quarkonia through HQ free energies and Euclidean correlation functions~\cite{Petreczky:2005nh,Petreczky:2005zy,Petreczky:2008px,Ding:2012sp}, which constrain calculations of spectral functions that can serve as an interface to phenomenological applications~\cite{Rapp:2009my,Mocsy:2013syh,He:2022ywp}.  

Open heavy-flavor (HF) particles have emerged as an excellent probe of the transport properties of the QCD medium in URHICs~\cite{Prino:2016cni,Rapp:2018qla,Dong:2019unq}. Produced in initial hard processes, low-momentum heavy quarks exert a Brownian motion through the QGP characterized by a spatial diffusion coefficient, hadronize in different HF hadrons and subsequently are further transported through the hadronic medium. The large HQ mass implies the dominance of elastic interactions with small energy transfer amenable to potential approximations, and the final HF baryon spectra carry a memory of their interaction history due to a thermalization time being comparable or larger than the fireball lifetime. 

The present work builds on previous efforts to develop a quantum many-body theory to describe the spectral and transport properties of open and hidden HF particles in a strongly coupled QGP~\cite{Cabrera:2006wh,Riek:2010fk}, including the 1- and 2-body 
Green's functions of thermal partons for obtaining the equation of state (EoS) in a selfconsistent Brueckner scheme~\cite{Mannarelli:2005pz,Liu:2018syc}. 
The basic ingredient to this framework is the 2-body interaction kernel for the in-medium $T$-matrix for which we employ an ansatz using a Cornell potential, whose temperature corrections are constrained by lQCD data for the HQ free energy. 
A salient feature of this approach is that it recovers basic features of vacuum spectroscopy (such as masses of quarkonia, $D$ and $B$ mesons, and non-Goldstone light hadrons), providing a baseline for the calculation of medium effects. In the spirit of a $1/M_Q$ expansion, spin-orbit and spin-spin interactions were not included thus far. In the present paper, we take the next step by including the latter by benchmarking them against the hyper-/fine mass splittings of quarkonia in vacuum.

Our study raises the question of the Lorentz structure of the confining potential. Historically, a default assumption of a purely scalar interaction has been employed ~\cite{Mur:1992xv,Lucha:1991vn}, implying a vanishing long-range magnetic contribution that nevertheless could reproduce the empirical fine structure for heavy quarkonium~\cite{Buchmuller:1981fr}. However, studies of the Wilson loop suggest that the confining potential cannot be a purely scalar kernel~\cite{Brambilla:1996aq,Brambilla:1997kz}, and the latter also causes problems in constructing a stable vacuum of QCD~\cite{Szczepaniak:1996tk}. In the relativistic quark model~\cite{Ebert:1997nk,Ebert:2002pp} a mixing of scalar and vector structures in the confining potential has been found to yield a quarkonium spectroscopy in good overall agreement with experimental data. The approach we employ in the present paper is close in spirit to these works, \ie, we will incorporate the possibility of a mixed confining Lorentz structure in the $T$-matrix kernel with the goal of improving the description of the observed the hyper/fine splittings in the vacuum quarkonium spectroscopy; the pertinent relativistic corrections will turn out to have significant ramifications for the HQ diffusion coefficient. 

This article is organized as follows. In Sec.~\ref{sec_TM}, we briefly recollect the main elements of the thermodynamic $T$-matrix approach. In Sec.~\ref{sec_Vvac} we implement spin-dependent interactions as well as a vector component of the confining force into the potential. In Sec.~\ref{sec_spectroscopy} we compute heavy-quarkonium spectral functions from the $T$-matrix and discuss the charmonium and bottomonium spectroscopy in vacuum. In Sec.~\ref{sec_qgp} we lay out our constraints on the in-medium corrections to the potential using lQCD data for static HQ free energies (Sec.~\ref{ssec_FQQ}) and the QGP equation of state (Sec.~\ref{ssec_eos}), and  discuss the pertinent numerical results (Sec.~\ref{ssec_self-con}). In Sec.~\ref{sec_transport} we outline the calculation of the HQ transport coefficients and highlight the implications of the vector component in the confining interaction on the numerical results for charm quarks. We summarize and conclude in Sec.~\ref{sec_concl}.

\section{T-matrix Approach}
\label{sec_TM} 
\begin{figure}[htbp]
\begin{minipage}[b]{1.0\linewidth}
\centering
\includegraphics[width=1.0\textwidth]{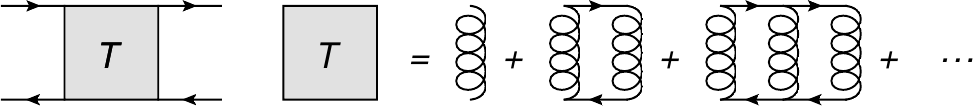}
\end{minipage}
\caption{$T$-matrices resummation for ladder diagrams.} 
\label{fig_Tm}
\end{figure}

The thermodynamic $T$-matrix is a 2-particle irreducible (PI) quantum many-body scheme that selfconsistently solves the 1- and 2-body Green's functions and is thus suitable for strongly interacting systems. In Refs.~\cite{Mannarelli:2005pz,Riek:2010fk} it has been initially developed to study the properties of HF particles in the QGP allowing for a reduction of the 4-dimensional (4D) Bethe-Salpeter 2-body scattering equation to a 3D one which allows for tractable numerical solutions. Subsequently, it has also been extended to the light-parton sector~\cite{Liu:2017qah}, based on the notion that the effective masses of the QGP's constituents are typically large compared to temperatures not too far above the pseudocritcal one
of $T_{\rm pc}\simeq 160$~MeV. The starting point can be formulated in terms of an effective Hamiltonian with a relativistic potential, 
\begin{eqnarray}
H= & \sum \varepsilon_i(\mathbf{p}) \psi_i^{\dagger}(\mathbf{p}) \psi_i(\mathbf{p})+\frac{1}{2} \psi_i^{\dagger}\left(\frac{\mathbf{P}}{2}-\mathbf{p}\right) \nonumber \\
& \times \psi_j^{\dagger}\left(\frac{\mathbf{P}}{2}+\mathbf{p}\right) V_{i j}^a \psi_j\left(\frac{\mathbf{P}}{2}+\mathbf{p}^{\prime}\right) \psi_i\left(\frac{\mathbf{P}}{2}-\mathbf{p}^{\prime}\right)
\label{eq_H}
\end{eqnarray}
which emphasizes the implementation of unitarity through resummations of the propagators (also referred to as a Dyson-Schwinger set-up). Here, 
$\mathbf{p}$ and $\mathbf{p'}$ denote the relative momentum of the incoming and outgoing states, and $\mathbf{P}$ the total momentum of the two-body system. Furthermore, $\varepsilon_i(\mathbf{p})=\sqrt{M_{i}^{2}+\mathbf{p}^{2}}$ is the dispersion relation of a parton with mass $M_i$, and the $V_{ij}^{a}$  are the potentials between particles $i$ and $j$ in a color channel $a$. The summation includes momentum, spin, color and flavor for quarks and gluons.
The infinite series of ladder diagrams generated by the Hamiltonian in Eq.~(\ref{eq_H}) straightforwardly results in the $T$-matrix equation, depicted in Fig.~\ref{fig_Tm}.
In the center-of-mass (CM) frame, one has 
\begin{eqnarray}
\ T_{ij}^{a}\left ( z,\mathbf{p},\mathbf{{p}'} \right )&=&V_{ij}^{a}\left ( \mathbf{p},\mathbf{{p}'} \right ) +\int_{-\infty}^{\infty}\frac{d^{3}\mathbf{k}}{\left ( 2\pi  \right )^{3}}V_{ij}^{a}\left ( \mathbf{p},\mathbf{k} \right )\nonumber \\
&&\times G_{ij}^{0}\left ( z,\mathbf{k} \right )T_{ij}^{a}\left ( z,\mathbf{k},\mathbf{{p}'} \right ) \ ,
\label{eq_3DTM}
\end{eqnarray}
where $G_{ij}^{0} $ is the two-body propagator, $z=E\pm i\epsilon$ the analytical energy variable, and $\mathbf{p}$ and $\mathbf{p'}$ are the incoming and outgoing 3-momenta in the CM frame, respectively. The reduction scheme from 4D to 3D is not unique~\cite{Brockmann:1996xy} but its specific choice has minor impact on the results; we choose the Thompson scheme following our previous studies~\cite{Riek:2010fk,Liu:2017qah}. In this scheme, the two-body propagator in spectral representation can be written as
\begin{eqnarray}
G_{i j}^{0}(z, \mathbf{k})&=& \int_{-\infty}^{\infty} d \omega_{1} d \omega_{2} \frac{\left[1 \pm n_{i}\left(\omega_{1}\right) \pm n_{j}\left(\omega_{2}\right)\right]}{z-\omega_{1}-\omega_{2}}\nonumber \\
&&\times \rho_{i}\left(\omega_{1}, \mathbf{k}\right) \rho_{j}\left(\omega_{2}, \mathbf{k}\right) \ , 
\label{eq_2bodyprop}
\end{eqnarray}
with the single-particle propagator
\begin{equation}
G_{i}(z)=\frac{1}{\left[G_{i}^{0}(z, k)\right]^{-1}-\Sigma_{i}(z, k)}=\frac{1}{z-\varepsilon_{i}(k)-\Sigma_{i}(z, k)}
\label{eq_1bodyprop}
\end{equation}
and the single-particle spectral function
\begin{equation}
\rho_{i}\left(\omega_{,} \mathbf{k}\right)=-\frac{1}{\pi} \operatorname{Im} G_{i}(\omega+i \epsilon) \ .
\label{eq_1bodyspec}
\end{equation}
 The ± signs in Eq.~(\ref{eq_2bodyprop}) correspond to bosons (upper) or fermions (lower)\footnote{The convention that upper/lower signs denote bosons/fermions is applied throughout this work.}, and $n_i$ is the Bose or Fermi distribution function for parton $i$. 
 In quasi-particle approximation Eq.~(\ref{eq_2bodyprop}) reduces to\footnote{This differs from Refs.~\cite{Brockmann:1996xy,Riek:2010fk} by a factor of $m_{ij}(\mathbf{k})=\frac{M_{i}M_{j}}{\varepsilon_i(\mathbf{k})\varepsilon_j(\mathbf{k})}$; here we keep the convention of Ref.~\cite{Liu:2017qah} where $m_{ij}(\mathbf{k})$ is absorbed into the relativistic corrections to the potential which will be elaborated in Sec.~\ref{subsec_VLS}.} 
\begin{eqnarray}
G_{ij}^{0}(z,\mathbf{k})&=& \frac{1}{z-\varepsilon_i(\mathbf{k})-\varepsilon_j(\mathbf{k})- \Sigma_{i}(\mathbf{k})-\Sigma_{j}(\mathbf{k})} \ .
\label{eq_2vacbodyprop}
\end{eqnarray}
The single-particle selfenergies in the QGP, $\Sigma_{i}(\mathbf{k})$, are obtained by closing the $T$-matrix with an in-medium single-parton propagator from the heat bath; its spectral representation is
\begin{eqnarray}
\Sigma_{i}\left(z, \mathbf{p}_{1}\right) &=&\frac{1}{d_{i}} \int \frac{d^{3} \mathbf{p}_{2}}{(2 \pi)^{3}} \int_{-\infty}^{\infty} d \omega_{2} \frac{d E}{\pi} \frac{-1}{z+\omega_{2}-E}  \nonumber \\
&& \times \sum_{a, j} d_{s}^{i j} d_{a}^{i j} \operatorname{Im} T_{i j}^{a}\left(E, \mathbf{p}_{1}, \mathbf{p}_{2} \mid \mathbf{p}_{1}, \mathbf{p}_{2}\right)\nonumber \\
&& \times \rho_{j}\left(\omega_{2}, \mathbf{p}_{2}\right)\left[n_{j}\left(\omega_{2}\right) \mp n_{i j}(E)\right],
\label{eq_selfenergy}
\end{eqnarray}
with $T\left(E, \mathbf{p}_{1}, \mathbf{p}_{2} \mid \mathbf{p}_{1}, \mathbf{p}_{2}\right)$ the forward-scattering $T$-matrix, i.e., $\mathbf{p}_{1}^{\prime}=\mathbf{p}_{1}$ and $\mathbf{p}_{2}^{\prime}=\mathbf{p}_{2}$, where $\mathbf{p}_{1,2}$ and $\mathbf{p}_{1,2}^{\prime}$ are the incoming and outgoing momenta for particle 1 and 2 respectively, defined in the thermal frame. The $n_{ij}$ refers to the thermal distribution for the two-body state $ij$, while $\mp$  refers to the bosonic/fermionic single-parton state $i$. The $d_{a, s}^{i j}$ are color and spin degeneracies of the two-body system, and $d_{i}$ is the spin-color degeneracy of the single parton $i$. We also need to add the purely real thermal Fock term~\cite{fetter2012quantum},
\begin{eqnarray}
\Sigma_{i}\left(\mathbf{p}_{1}\right) &=& \mp \int \frac{d^{3} \mathbf{p}_{2}}{(2 \pi)^{3}} \int_{-\infty}^{\infty} d \omega_{2} V_{i \bar{i}}^{a=1}\left(\mathbf{p}_{1}-\mathbf{p}_{2}\right) \rho_{i}\left(\omega_{2}, \mathbf{p}_{2}\right) \nonumber \\
&&\times n_{i}\left(\omega_{2}\right) \ ,
\label{eq_fockselfenergy}
\end{eqnarray}
which is not part of the selfenergy in Eq.~(\ref{eq_selfenergy}). The $V_{i \bar{i}}^{a=1}$ refers to the color-singlet potential between particle and antiparticle. The selfenergy can be solved selfconsistently by iterating Eqs.~(\ref{eq_3DTM}), (\ref{eq_selfenergy}) and (\ref{eq_fockselfenergy}) numerically.  In doing so, the $T$-matrix in the thermal frame, $T_{i j}^{a}\left(\omega_{1}+\omega_{2}, \mathbf{p}_{1}, \mathbf{p}_{2} \mid \mathbf{p}_{1}^{\prime}, \mathbf{p}_{2}^{\prime}\right)$ needs to be transformed into the the CM frame, $T_{i j}^{a}\left(E_{\mathrm{cm}}, p_{\mathrm{cm}}, p_{\mathrm{cm}}^{\prime}, \cos(\theta_{\mathrm{cm}})\right)$. This is accomplished by
\begin{eqnarray}
E_{\mathrm{cm}} &=&\sqrt{\left(\omega_{1}+\omega_{2}\right)^{2}-\left(\mathbf{p}_{1}+\mathbf{p}_{2}\right)^{2}} \nonumber \\
s_{\mathrm{on}} &=&\left(\varepsilon_{1}\left(\mathbf{p}_{1}\right)+\varepsilon_{2}\left(\mathbf{p}_{2}\right)\right)^{2}-\left(\mathbf{p}_{1}+\mathbf{p}_{2}\right)^{2} \nonumber \\
p_{\mathrm{cm}} &=&\sqrt{\frac{\left(s_{\mathrm{on}}-M_{i}^{2}-M_{j}^{2}\right)^{2}-4 M_{i}^{2} M_{j}^{2}}{4 s_{\mathrm{on}}}} \nonumber \\
\cos \left(\theta_{\mathrm{cm}}\right) &=&\frac{\mathbf{p}_{\mathrm{cm}} \cdot \mathbf{p}_{\mathrm{cm}}^{\prime}}{p_{\mathrm{cm}} p_{\mathrm{cm}}^{\prime}},
\label{eq_trans}
\end{eqnarray}
where $\cos(\theta_{\mathrm{cm}})$ is the angle between the incoming and outgoing momenta in the CM frame, and $p_{\mathrm{cm}}^{\prime}$ can be obtained by substituting $s_{\mathrm{on}}(\mathbf{p}_{1},\mathbf{p}_{2})$ with $s_{\mathrm{on}}(\mathbf{p}_{1}^{\prime},\mathbf{p}_{2}^{\prime})$. As discussed in Ref.~\cite{Liu:2017qah}, the reason for using the on-shell value, $s_{on}$, for $p_{\mathrm{cm}}$ is to preserve the analytical properties of the $T$-matrix after the transformation into the CM frame.

The 3D $T$-matrix integral equation can be further reduced to a 1D one by applying the partial-wave expansion in the CM frame 
(from hereon the subscript ``cm" is suppressed for simplicity),
\begin{equation}
X(\mathbf{p},\mathbf{p'})=4\pi \sum_{L}(2L+1)X^{L}(p,p')P_{L}[\cos(\theta ) ] \ ,
\label{eq_partial expansion}
\end{equation}
where $X$ denotes $V$ or $T$, $L$  the angular-momentum quantum number, and $p$ and $p'$ are the moduli of $\mathbf{p}$ and $\mathbf{p'}$. The 1D $T$-matrix equation then takes the form 
\begin{eqnarray}
\ T_{ij}^{L,a} ( z,p,p')&=&V_{ij}^{{L,a}} (p,p') +\frac{2}{\pi } \int_{-\infty}^{\infty}k^{2}dk V_{ij}^{{L,a}} (p,k)\nonumber \\
&&\times G_{ij}^{0} (z,k)T_{ij}^{L,a} ( z,k,p').
\label{eq_1DTM}
\end{eqnarray}
Equation~(\ref{eq_1DTM}) can be solved by discretizing the momenta to convert it into a matrix equation and solve it by matrix inversion.

\section{Two-Body Potentials in Vacuum}
\label{sec_Vvac} 

In this section we first discuss the static potentials in Sec.~\ref{subsec_Vstatic}, then introduce the relativistic corrections to the potential between particles $i$ and $j$, and construct the confining potential with mixed Lorentz structures in Sec.~\ref{subsec_VLS}. The potential is generalized to different color channels at the end of this section. For simplicity,  we suppress the color factors indices until the end of this section.

\subsection{Static Potential}
\label{subsec_Vstatic} 

\begin{figure}[htbp]
\begin{minipage}[b]{1.0\linewidth}
\centering
\includegraphics[width=1.0\textwidth]{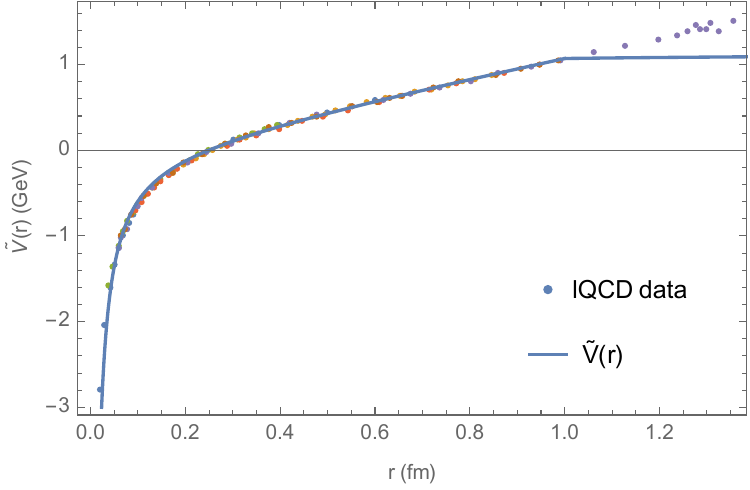}
\end{minipage}
\caption{The fitted vacuum potential versus the lQCD data. The blue line denotes the fitted vacuum potential, the colored dots the lQCD data from Refs.~\cite{Cheng:2007jq,Petreczky:2010yn,Mocsy:2013syh,Bazavov:2014kva,HotQCD:2014kol,Bazavov:2018wmo}.} 
\label{fig_Vvac}
\end{figure}

The kernel of the $T$-matrix equation (\ref{eq_3DTM}) is based on the Cornell potential, with a color-Coulomb potential, $V_{C}$, plus a confining potential (``string" term), $V_{S}$. In coordinate space the common ansatz is
\begin{equation}
\widetilde{V}(r)={\widetilde{V}}_{\mathcal{C}}(r)+{\widetilde{V}_{\mathcal{S}}}(r)=-\frac{4}{3}\frac{\alpha_s }{r}+\sigma r \ ,
\label{eq_VCornellr}
\end{equation}
where $\alpha_s$ and $\sigma$ are the perturbative coupling constant and nonperturbative string tension, respectively. To obtain the momentum-space potentials, ${V}_{\mathcal{C}/\mathcal{S}}(\mathbf{k})$, depending on the momentum transfer $\mathbf{k}=\mathbf{p-p'}$, we use the subtracted quantities ${V}_{\mathcal{C}/\mathcal{S}}(r)={\widetilde{V}}_{\mathcal{C}/\mathcal{S}}(r)-{\widetilde{V}}_{\mathcal{C}/\mathcal{S}}(\infty)$ to ensure the convergence of the Fourier transforms. A running coupling is implemented in the Coulomb potential for off-shell scattering in momentum space as~\cite{Riek:2010fk} $F_{run}(p,p')=\ln \left[\frac{\Delta^{2}}{\Lambda^{2}}\right] / \ln \left[\frac{(p-p')^{2}+\Delta^{2}}{\Lambda^{2}}\right]$. For the confining potential, we enforce a flat potential above a string breaking scale of about $r_{SB}=1$\,fm to account for string breaking. We employ the same potential parameters as in previous studies~\cite{Liu:2017qah}, i.e., $\alpha_s=0.27$, $\sigma=0.225$ $\textup{GeV}^2$, $\Delta=1$ GeV and $\Lambda=0.2$ GeV, which are fitted to the lQCD data of the vacuum free energy~\cite{Cheng:2007jq,Petreczky:2010yn,Mocsy:2013syh,Bazavov:2014kva,HotQCD:2014kol,Bazavov:2018wmo} as shown in Fig.~\ref{fig_Vvac}, noting that in vacuum the color-singlet free energy is identical to the potential (as there is no entropy term).

\subsection{Relativistic Corrections and Spin-Dependent Interactions}
\label{subsec_VLS} 
Relativistic effects in the one-gluon exchange amplitude are well known, containing spin-independent and spin-dependent ones. For the pertinent vector potential (denoted as $V^{vec}$), the spin-independent correction amounts to multiplying the momentum-space potential by a factor
\begin{eqnarray}
\mathcal{R}_{i j}&=&\sqrt{\frac{1}{m_{ij}(p)}}\sqrt{1+\frac{p^{2}}{\varepsilon _{i}(p)\varepsilon _{j}(p)}}\nonumber \\
&&\times\sqrt{\frac{1}{m_{ij}(p')}}\sqrt{1+\frac{p'^{2}}{\varepsilon _{i}(p')\varepsilon _{j}(p')}} \ ,
\label{eq_R correction}
\end{eqnarray}
which is known as the Breit correction, representing magnetic effects. For scalar potentials, denoted as $V^{sca}$, no relativistic correction arises to leading order in $1/M_Q$, see Ref.~\cite{Riek:2010fk}. We write the total spin-independent potential in momentum space as
\begin{eqnarray}
V_{i j}\left(\mathbf{p}, \mathbf{p}^{\prime}\right)&=&\mathcal{R}_{i j} V^{vec}\left(\mathbf{p}-\mathbf{p}^{\prime}\right)+V^{sca}\left(\mathbf{p}-\mathbf{p}^{\prime}\right). 
\label{eq_V correction}
\end{eqnarray}
To implement spin-dependent interactions, including spin-orbit ($V^{LS}$), spin-spin ($V^{SS}$) and tensor ($V^{T}$) channels, we follow Ref.~\cite{Lucha:1995zv} where the detailed procedure to derive the Fermi-Breit Hamiltonian is laid out. The pertinent corrections for vector and scalar potentials between two partons with equal masses ($M_i=M_j\equiv M$) in coordinate space are given by
\begin{eqnarray}
{V^{LS}}(r) &=&\frac{1}{2M^{2}r}\left \langle \mathbf{L\cdot S} \right \rangle \left (3\frac{d}{dr}V^{vec}(r)-\frac{d}{dr}V^{sca}(r)\right),\nonumber\\
{V^{SS}}(r) &=&\frac{2}{3M^{2}}\left \langle \mathbf{S_1\cdot S_2} \right \rangle \Delta V^{vec}(r),\nonumber\\
{V}^{T}(r) &=&\frac{1}{12M^{2}}\left \langle S_{12} \right \rangle \left (\frac{1}{r}\frac{d}{dr}V^{vec}(r)-\frac{d^2}{dr^2}V^{vec}(r)\right) \ , \nonumber\\ 
&
\label{eq_VLSSST}
\end{eqnarray}
where $\Delta\equiv\nabla^2$ in the $SS$ interaction is the Laplace operator.
Note that the scalar interactions do not contribute to the $SS$ and $T$ corrections. We note that the vector potential, $V^{vec}$, in the spin-dependent potentials above do not receive the spin-independent Breit correction, $\mathcal{R}$, introduced in Eq.~(\ref{eq_R correction}).
Following Ref.~\cite{Hong:2022sht}, we smear the Dirac delta function $\delta$ in the $SS$ part by a Gaussian, $\tilde{\delta}(r)=\left(\frac{b}{\sqrt{\pi}}\right)^3 e^{-b^2 r^2}$, to avoid the singularity. We take $b=10$ in this work and have checked that for $b>$10 the $SS$ interaction saturates in the quarkonium spectroscopy.

The expectation values take the standard form (with $L$, $S$ and $J$ denoting the orbital, spin and total angular-momentum quantum numbers, respectively): $\left \langle \mathbf{L\cdot S} \right \rangle=\frac{1}{2}[J(J+1)-L(L+1)-S(S+1)]$, $\left \langle \mathbf{S_1\cdot S_2} \right \rangle=\frac{1}{2}[S(S+1)-\frac{3}{2}]$, and $\left\langle S_{12}\right\rangle=\frac{4}{(2 L+3)(2 L-1)}\left[S(S+1)J(J+1)-\frac{3}{2}\left \langle \mathbf{L\cdot S} \right \rangle-3(\left \langle \mathbf{L\cdot S} \right \rangle)^{2}\right]$ for $L\neq0$ and $S=1$, but $\left\langle S_{12}\right\rangle$ vanishes for either $L=0$ or $S=0$. 
The total potential (with relativistic corrections) between, \eg, a heavy quark and anti-quark in momentum-space 
reads
\begin{eqnarray}
&V_{Q\bar{Q}}\left(\mathbf{p}, \mathbf{p}^{\prime}\right)=\mathcal{R}_{Q\bar{Q}} V^{vec}\left(\mathbf{p}-\mathbf{p}^{\prime}\right)+ V^{sca}(\mathbf{p}-\mathbf{p'}) \nonumber\\
& \quad +V^{LS}(\mathbf{p}-\mathbf{p'})+V^{SS}(\mathbf{p}-\mathbf{p'})+V^{T}(\mathbf{p}-\mathbf{p'}) \ , 
\label{eq_V0}
\end{eqnarray}
where the spin-dependent terms in momentum-space are obtained through Fourier transform, $V^{a}(\mathbf{k}=\mathbf{p}-\mathbf{p'})=\int d^{3}\mathbf{r}e^{-i\mathbf{k}\cdot \mathbf{r}}V^{a}(\mathbf{r})$ with $a=LS,SS,T$.
We absorb $m_{ij}(\mathbf{k})$ in the two-body propagator into the relativistic corrections for the potentials to keep the same convention as in Ref.~\cite{Liu:2017qah}, thus Eqs.~(\ref{eq_V correction}) become
\begin{eqnarray}
&V_{i j}\left(\mathbf{p}, \mathbf{p}^{\prime}\right)\rightarrow \sqrt{{m_{ij}(p)}}\sqrt{{m_{ij}(p')}} V_{ij}\left(\mathbf{p}, \mathbf{p}^{\prime}\right) \nonumber\\
& \qquad =\mathcal{R}_{i j}^{vec} V^{vec}\left(\mathbf{p}-\mathbf{p}^{\prime}\right)+\mathcal{R}_{i j}^{sca}V^{sca}\left(\mathbf{p}-\mathbf{p}^{\prime}\right), 
\label{eq_V correction new}
\end{eqnarray}
with
\begin{eqnarray}
\mathcal{R}_{i j}^{vec}&\equiv &\sqrt{{m_{ij}(p)}}\sqrt{{m_{ij}(p')}}\mathcal{R}_{i j}
\nonumber \\
&=&\sqrt{1+\frac{p^{2}}{\varepsilon _{i}(p)\varepsilon_{j}(p)}}\sqrt{1+\frac{p'^{2}}{\varepsilon _{i}(p')\varepsilon _{j}(p')}},\nonumber \\
\mathcal{R}_{i j}^{sca}&\equiv &\sqrt{{m_{ij}(p)}}\sqrt{{m_{ij}(p')}}\nonumber \\
&=&\sqrt{\frac{M_{i}M_{j}}{\varepsilon_i(p)\varepsilon_j(p)}}\sqrt{\frac{M_{i}M_{j}}{\varepsilon_i(p')\varepsilon_j(p')}} \ .
\label{eq_R correction new}
\end{eqnarray}
and Eq.~(\ref{eq_V0}) becomes 
\begin{eqnarray}
&V_{Q\bar{Q}}\left(\mathbf{p}, \mathbf{p}^{\prime}\right)\rightarrow \sqrt{{m_{ij}(p)}}\sqrt{{m_{ij}(p')}} V_{Q\bar{Q}}\left(\mathbf{p}, \mathbf{p}^{\prime}\right) \qquad \qquad
\nonumber\\
& =\mathcal{R}_{Q\bar{Q}}^{vec} V^{vec}\left(\mathbf{p}-\mathbf{p}^{\prime}\right)+\mathcal{R}_{Q\bar{Q}}^{sca}V^{sca}\left(\mathbf{p}-\mathbf{p}^{\prime}\right) \qquad
\nonumber\\
& \qquad +\mathcal{R}_{Q\bar{Q}}^{spin}[V^{LS}(\mathbf{p}-\mathbf{p'})+V^{SS}(\mathbf{p}-\mathbf{p'})+V^{T}(\mathbf{p}-\mathbf{p'})] \nonumber\\
\label{eq_V}
\end{eqnarray}
with
\begin{eqnarray}
\mathcal{R}_{i j}^{spin}&\equiv &\sqrt{{m_{ij}(p)}}\sqrt{{m_{ij}(p')}}\nonumber \\
&=&\sqrt{\frac{M_{i}M_{j}}{\varepsilon_i(p)\varepsilon_j(p)}}\sqrt{\frac{M_{i}M_{j}}{\varepsilon_i(p')\varepsilon_j(p')}} \ .
\label{eq_R correction SS}
\end{eqnarray}
The Lorentz structure for Coulomb potential is entirely vector, and a common assumption for the confining one is to be entirely scalar, \ie,  $V^{vec}=V_\mathcal{C}$ and $V^{sca}=V_\mathcal{S}$. As was mentioned in the introduction, there are reasons to believe that the confining potential is not a purely scalar one but a mixture of vector and scalar Lorentz structures, \ie, 
$V^{vec}=V_\mathcal{C}+(1-\chi)V_\mathcal{S}$ and $V^{sca}=\chi V_\mathcal{S}$. The key parameter is the mixing coefficient, $\chi$, defined such that for $\chi=1$ the interaction reduces to the case with a purely scalar confining potential, while values below one characterize a vector admixture.

The potentials in the various color channels, $V_{ij}^{a}\left ( \mathbf{p},\mathbf{{p}'} \right )$, are obtained by the substitutions $V_{\mathcal{C}}\left(\mathbf{p}-\mathbf{p}^{\prime}\right) \rightarrow \mathcal{F}_{a}^{\mathcal{C}} V_{\mathcal{C}}\left(\mathbf{p}-\mathbf{p}^{\prime}\right)$ and $V_{\mathcal{S}}\left(\mathbf{p}-\mathbf{p}^{\prime}\right) \rightarrow \mathcal{F}_{a}^{\mathcal{S}} V_{\mathcal{S}}\left(\mathbf{p}-\mathbf{p}^{\prime}\right)$. For the Coulomb interaction, $ \mathcal{F}_{a}^{\mathcal{C}}$ are the standard Casimir coefficients listed in Tab. ~\ref{tab_color} (together with the pertinent degeneracy factors), and we take the absolute values of the Casimir coefficients for the string interaction, $ \mathcal{F}_{a}^{\mathcal{S}}$, to ensure a positive definite string tension~\cite{Liu:2017qah}.  
\begin{table}[ht]
\tabcolsep 0pt \caption{\label{tab_color}Casimir and degeneracy factors for different color channels (Casimir factor, degeneracy).}
\setlength{\tabcolsep}{11pt}
\begin{tabular}{lccc}
\hline \hline$q q$ & $q \bar{q}$ & $(q / \bar{q}) g$ & $g g$ \\
\hline$(1 / 2,3)$ & $(1,1)$ & $(9 / 8,3)$ & $(9 / 4,1)$ \\
$(-1 / 4,6)$ & $(-1 / 8,8)$ & $(3 / 8,6)$ & $(9 / 8,16)$ \\
& & $(-3 / 8,15)$ & $(-3 / 4,27)$ \\
\hline \hline
\end{tabular}
\end{table}

The parton masses are also related to the potential introduced above. The constituent masses of the heavy quarks, $M_Q$, receives two contributions, the first one is calculated by the selfenergy from the color-singlet ($a=1$) potential (including the relativistic factors) and the second one is a ``bare mass", $M^0_{Q}$, which is associated with condensate contributions that we do not calculate explicitly in the present framework, 
\begin{eqnarray}
M_{Q}=-\frac{1}{2} \int \frac{d^{3} \mathbf{p}}{(2 \pi)^{3}} V_{Q  \bar{Q}}^{a=1}(\mathbf{p})+M^0_{Q},
\label{eq_constituentM}
\end{eqnarray}

\section{Heavy-Quarkonium Spectroscopy in Vacuum}
\label{sec_spectroscopy} 
In this section we introduce the correlation and spectral functions including their non-relativistic classifications in angular momentum 
(Sec.~\ref{ssec_corr-fcts}) and discuss our fits to the vacuum spectra including the spin-related interactions (Sec.~\ref{ssec_vac-fit}).

\subsection{Correlation and Spectral Functions}
\label{ssec_corr-fcts}
To evaluate the quarkonia spectra in both charm and bottom sectors, we compute the quark-antiquark spectral functions for different mesonic quantum-numbers channels using the pertinent  $T$-matrices as described in Sec.~\ref{sec_TM}. The bound-state masses are then determined from the peak values of corresponding mesonic spectral functions. In the vacuum, we introduce a small width in the single-quark propagators which allows us to numerically resolve the bound-state mass while not distorting their masses. Since we only account for the $Q\bar Q$ channels (off-shell) couplings to intermediate two-meson states (\eg, $DD^*$ channels) are not accounted for, which could affect the masses near the $D\bar D$ threshold somewhat. 

Table~\ref{tab_LSJ} lists the $L$, $S$ and $J$ assignments in the scalar (S), pseudoscalar (PS), vector (V), axial-vector (AV) and tensor (T) mesonic channels. In practice, a cutoff $r_{c}=0.01$ fm is introduced in the Fourier transform for the spin-dependent potentials to avoid ultraviolet divergences; we have checked that the results are not sensitive to variations in  $r_{c}$ by $\pm 50\%$.
\begin{table}[ht]
\tabcolsep 0pt \caption{\label{tab_LSJ}Non-relativistic classification of angular-momentum quantum numbers in different mesonic channels.}
\setlength{\tabcolsep}{20pt}
\begin{center}
\def\temptablewidth{0.8\textwidth}
\begin{tabular}{c c c c c c}
\hline
\hline
  Channels        & $L$      & $S$   & $J$       \\
   \hline
   S        & $1$      &$1$  & $0$  \\

  PS       & $0$       & $0$ & $0$  \\

   V        & $0 $     &  $1$ & $1$  \\

   AV$_{1}$        & $1$   & $0$ & $1$    \\
   
   AV$_{2}$        & $1$   & $1$ & $1$    \\
   
   T        & $1$   & $1$ & $2$    \\
   \hline
\hline
\end{tabular}
\end{center}
\end{table}

\begin{figure}[htbp]
\begin{minipage}[b]{1.0\linewidth}
\centering
\includegraphics[width=1.0\textwidth]{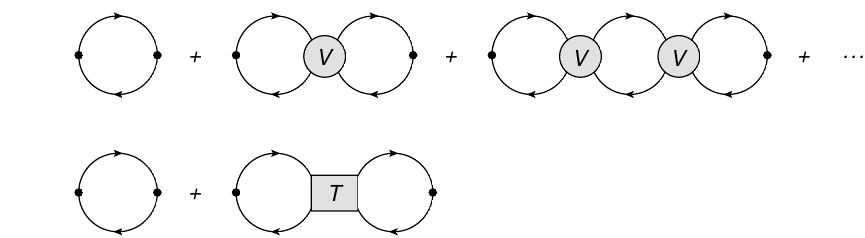}
\end{minipage}
\caption{Diagrammatic representation of the $Q\bar{Q}$ correlation function. The dots denote meson current operators $\Gamma_M$.} 
\label{fig_G}
\end{figure}
The spectral functions are obtained form the correlation functions, $G$, of the meson currents. The latter are obtained by closing the two incoming and outgoing legs of $T$-matrix (plus a non-interacting contribution) with the corresponding projection operator for the different quarkonium channels, see Fig.~\ref{fig_G}. Writing
\begin{equation}
G=G_0+\Delta G \ , 
\label{eq_G}
\eeq
the non-interacting part of correlation function in the CM frame is given by
\begin{eqnarray}
\ G_{0}(E,T)&=&N_{f}N_{c}\int \frac{d^3p}{(2\pi )^3}\mathcal{R}_{Q\bar{Q}}^{sca}\nonumber \\
&&\times \mathrm{Tr}\left \{ \Gamma_\alpha\Lambda _+(\mathbf{p})\Gamma_\alpha\Lambda _-(-\mathbf{p}) \right \}G_{Q\bar{Q}}^0(E,p),\nonumber \\
\label{eq_G0}
\end{eqnarray}
where $\Lambda _\pm(\mathbf{p})=[\varepsilon _Q(p)\gamma^0-(\mathbf{p}\cdot \bm{\gamma})\pm M_Q]/2M_Q$ are the positive/negative energy projectors for quark and antiquark, respectively, and $\Gamma_\alpha \in (1,i\gamma_5,\gamma^\mu,\gamma^\mu \gamma_5,\frac{i}{2}[\gamma ^{\mu },\gamma ^{\nu }] )$ the vertex operators for the mesonic S, PS, V, AV and T channels, respectively; the pertinent traces are listed in Tab.~\ref{tab_tr-G0}. Furthermore,  $N_f=1$ and $N_c=3$ are the numbers of flavor and color for the heavy quark.

\begin{table}[ht]
\tabcolsep 0pt \caption{Values for the trace coefficients in different mesonic channels for the non-interacting part, $G_0$, of the correlation functions.}
\setlength{\tabcolsep}{40pt}
\begin{center}
\def\temptablewidth{0.8\textwidth}
\begin{tabular}{c c c c c c}
\hline
\hline
  $\Gamma_\alpha$        & $G_0$ trace             \\
   \hline
   S        & $2\frac{p^2}{M_Q^2}$         \\

  PS       & $2(1+\frac{p^2}{M_Q^2}) $         \\

   V        & $6+4\frac{p^2}{M_Q^2}$          \\

   AV        & $4\frac{p^2}{M_Q^2}$        \\
   
   T        & $4\frac{p^2}{M_Q^2}$        \\
   \hline
\hline
\end{tabular}
\end{center}
\label{tab_tr-G0}
\end{table}

The interacting part of correlation function in the CM frame is given by
\begin{eqnarray}
\Delta G(E,T)&=&\frac{N_fN_c}{8\pi ^4}\int dp p^2  G_{Q\bar{Q}}^0(E,p)\int dp' p'^2 G_{Q\bar{Q}}^0(E,p')\nonumber \\
&&\times \mathcal{R}_{Q\bar{Q}}^{sca} \mathcal{T}(\Gamma_\alpha;E,p,p'),
\end{eqnarray}
with the scattering amplitude
\begin{eqnarray}
\begin{aligned}
\mathcal{T}(\Gamma_\alpha;E,p,p')&=\int d(\cos \theta )\textup{Tr}(\Gamma _\alpha;p,p',\theta )T_{Q\bar{Q}}( E,\mathbf{p},\mathbf{{p}'} )\\
&=8\pi [a_0(p,p')T_{Q\bar{Q}}^{0}+a_1(p,p')T_{Q\bar{Q}}^{1}]
 \ .  
\end{aligned}
\end{eqnarray}
The $T$-matrix, $T_{Q\bar{Q}}^{L}$, is calculated from Eq.~(\ref{eq_1DTM}) with the interaction kernel in Eq.~(\ref{eq_V}). The $a_L$ denote the coefficients of the orbital-angular momentum, $L$, in the partial-wave expansion of the trace, 
\begin{eqnarray}
\textup{Tr}(\Gamma_\alpha;p,p',\theta )&=&\textup{Tr}\left \{ \Lambda _{+}(\mathbf{p})\Gamma _{\alpha}\Lambda _{-}(\mathbf{-p})\Lambda _{-}(\mathbf{-p'})\nonumber \right.\\ 
&&\left.\times \Gamma _{\alpha}\Lambda _{+}(\mathbf{p'}) \right \}
\nonumber \\
&=&a_0(p,p')P_0(\cos \theta )+a_1(p,p')P_1(\cos \theta ) \ ;
\nonumber\\
\label{eq_al}
\end{eqnarray}
they are listed in Tab.~\ref{tab_tr-G}. For the evaluation of the traces in the V, AV and T channels we focus on the spatial components. The correlation functions defined in Refs.~\cite{Cabrera:2006wh} and \cite{Riek:2010fk} do not have the $\mathcal{R}_{Q\bar{Q}}^{sca}$ factor due to the different definitions of potentials and two-body propagators, but they are equivalent to those in Refs.~\cite{Liu:2017qah} and in this work.

\begin{table}[!tbh]
\renewcommand\arraystretch{2}
\newcommand{\tabincell}[2]{\begin{tabular}{@{}#1@{}}#2\end{tabular}}
\tabcolsep 0pt \caption{Coefficients of orbital angular momentum (up to $L=1$) in different mesonic channels for the interacting part, $G$, of the correlation functions.}
\setlength{\tabcolsep}{8pt}
\begin{center}
\def\temptablewidth{0.8\textwidth}
\begin{tabular}{c c c c c c}
\hline
\hline
$\Gamma_\alpha$           & $a_0(p,p')$   & $a_1(p,p')$        \\
   \hline
   S               &  $-\frac{p^2p'^2}{M_Q^4}$      & $(1+\frac{\varepsilon(p)\varepsilon_Q(p')}{M_Q^2})\frac{pp'}{M_Q^2}$           \\

  PS              & \tabincell{c}{  $1+\frac{p^2+p'^2}{M_Q^2}$\\ $+\frac{\varepsilon_Q(p)\varepsilon_Q(p')}{M_Q^2}+\frac{p^2p'^2}{M_Q^4}$ }    & $-\frac{\varepsilon_Q(p)\varepsilon_Q(p')pp'}{M_Q^4}$        \\

   V               &   \tabincell{c}{$3(1+\frac{\varepsilon_Q(p)\varepsilon_Q(p')}{M_Q^2})$\\$+2\frac{p^2+p'^2}{M_Q^2}+\frac{4}{3}\frac{p^2p'^2}{M_Q^4}$ }     & $-(1+2\frac{\varepsilon_Q(p)\varepsilon_Q(p')}{M_Q^2})\frac{pp'}{M_Q^2}$           \\

   AV             &  $-\frac{4}{3}\frac{p^2p'^2}{M_Q^4}$   &  $2(1+\frac{\varepsilon_Q(p)\varepsilon_Q(p')}{M_Q^2})\frac{pp'}{M_Q^2}$       \\

   T             &  $-\frac{2}{3}\frac{p^2p'^2}{M_Q^4}$   &  $2(1+\frac{\varepsilon_Q(p)\varepsilon_Q(p')}{M_Q^2})\frac{pp'}{M_Q^2}$       \\
   \hline
\hline
\end{tabular}
\end{center}
\label{tab_tr-G}
\end{table}

At higher orders in the $1/M_Q$ expansion, the partial-wave expansion leads to a mixing between $S$- and $P$- wave components in a given meson channel. However, to keep with the (non-relativistic) classification of the meson channels with definite quantum numbers of $L$, $S$ and $J$, we terminate the expansion when the ``unnatural" partial waves admix~\cite{Cabrera:2006wh}. 
The pertinent leading orders are collected in Tab.~\ref{tab_tr-LO}.
\begin{table}[ht]
\tabcolsep 0pt \caption{Leading orders of  the angular-momentum coefficients, $a_L$, in evaluating the traces for the non-interacting (second column) and interacting (third and fourth columns) part of the different mesonic channels specified in the first column.}
\setlength{\tabcolsep}{3pt}
\begin{center}
\def\temptablewidth{0.8\textwidth}
\begin{tabular}{c c c c c c}
\hline
\hline
  $\Gamma_\alpha$        & $G_0$ trace      & $a_0(p,p')$   & $a_1(p,p')$       \\
   \hline
   S        & $2\frac{p^2}{M_Q^2}$      &$\mathcal{O}(p^4/M_{Q}^{4})$  & $-2\frac{pp'}{M_Q^2}+\mathcal{O}(p^4/M_{Q}^{4})$  \\

  PS       & $2+\mathcal{O}(p^2/M_{Q}^{2})$       & $-2+\mathcal{O}(p^2/M_{Q}^{2})$ & $\mathcal{O}(p^2/M_{Q}^{2})$  \\

   V        & $6 +\mathcal{O}(p^2/M_{Q}^{2}) $     &  $-6+\mathcal{O}(p^2/M_{Q}^{2})$ & $\mathcal{O}(p^2/M_{Q}^{2})$  \\

   AV        & $4\frac{p^2}{M_Q^2}$   & $\mathcal{O}(p^4/M_{Q}^{4})$ & $-4\frac{pp'}{M_Q^2}+\mathcal{O}(p^4/M_{Q}^{4})$    \\
   
   T        & $4\frac{p^2}{M_Q^2}$   & $\mathcal{O}(p^4/M_{Q}^{4})$ & $-4\frac{pp'}{M_Q^2}+\mathcal{O}(p^4/M_{Q}^{4})$    \\
   \hline
\hline
\end{tabular}
\end{center}
\label{tab_tr-LO}
\end{table}
From the correlation functions, the mesonic spectral functions follow from the imaginary part, 
\begin{equation}
\sigma_{\alpha}(E,T)= - \frac{1}{\pi}\operatorname{Im}G_\alpha(E+i\epsilon,T),
\label{eq_spec}
\end{equation}
where the subscript $\alpha$ denotes the different meson channels.

\subsection{Heavy-Quarkonium Spectra in Vacuum}
\label{ssec_vac-fit}
We are now in position to investigate the quantitative consequences of the spin-dependent interactions and the scalar-vector mixing effect in the confining potential on the charmonium and bottomonium spectroscopy in vacuum. 
In practice, we adopt a value for the HQ width of $\Gamma_Q=20$\,MeV which is small enough to not affect the vacuum masses, but large enough to allow for straightforward numerical computations and plotting. 
Our fit procedure is as follows: For a given mixing coefficient, $\chi$, we adjust the bare-quark masses $M_c^0$ ($M_b^0$) to find the best fit for all the masses of charmonium (bottomonium) states given by the Particle Data Group (PDG)~\cite{ParticleDataGroup:2018ovx} using a $\chi^2$ statistical test. In principle we could optimize the value for the mixing coefficient, $\chi$, by strictly minimizing the variance of the fit. 
However, in practice we found that $\chi=$0.6 already provides most of the improvement in the mass splittings compared to $\chi=1$, while for still smaller values the constituent quark masses become rather large producing uncomfortably large binding energies; in addition, a more precise evaluation of the quarkonium masses near the open heavy-flavor threshold would also require the inclusion of hadronic loop corrections.
In particular, we do not pursue here more extreme scenarios, as proposed, \eg, in Refs.~\cite{Ebert:1997nk,Ebert:2002pp} where optimized fits with $\chi=-1$ were found (implying a negative string tension for the scalar term, counteracted by a twice larger vector component).

We start by displaying the comparison of charmonium spectral functions between purely scalar ($\chi=1$) and mixed ($\chi=0.6$) confining potential for all states below the $D\bar D$ threshold in Fig.~\ref{fig_speccc} (we only plot the interacting parts of spectral functions since the free part does not affect the bound-state locations). The various peaks in each quantum-number channel are readily assigned as, S: $\chi_{c0} (1P)$; PS: $\eta_c(1S)$ and $\eta_c(2S)$; V: $J/\Psi (1S)$ and $\Psi(2S)$; AV$_1$: $h_c (1P)$;  AV$_2$: $\chi_{c1} (1P)$; T: $\chi_{c2} (1P)$. The masses extracted from the pole positions are listed in Tab.~\ref{tab_mass} and compared to the experimental values. The potential with mixing effect generates more attraction from the additional relativistic corrections (recall Eq.~(\ref{eq_R correction})), thus requiring a larger HQ mass for $\chi=0.6$ as quoted in Tab.~\ref{tab_mass}.  For $\chi=1$, neither the $S$- nor $P$-wave mass splittings are well reproduced; the results are much improved by introducing the mixing effect with $\chi=0.6$.
\begin{figure}[htbp]
\begin{minipage}[b]{1.0\linewidth}
\centering
\includegraphics[width=1.0\textwidth]{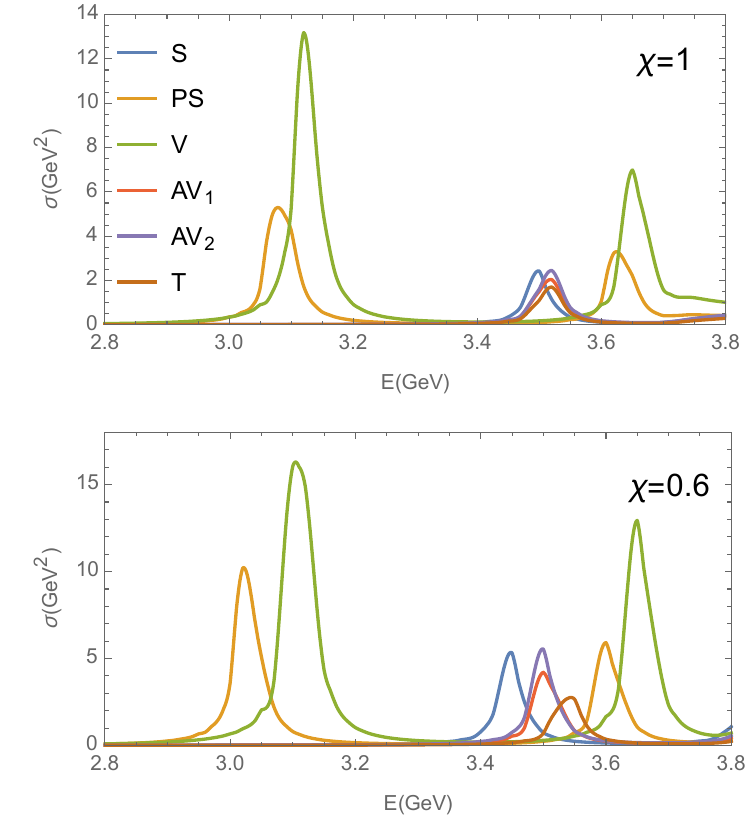}
\end{minipage}
\caption{The vacuum charmonium spectral functions (interacting parts) in S, PS, V, AV and T channels with mixing coefficient $\chi=1$ (upper panel) and $\chi=0.6$ (lower panel).} 
\label{fig_speccc}
\end{figure}
\begin{table}[ht]
\tabcolsep 0pt \caption{\label{tab_mass} Experimental and theoretical values for charmonium masses (in GeV) with the mixing coefficient $\chi=1$ and $\chi=0.6$; $M_c^0$ and $M_c$ are the bare and constituent charm-quark masses, respectively.}
\setlength{\tabcolsep}{6pt}
\begin{center}
\def\temptablewidth{0.8\textwidth}
\begin{tabular}{c c c c c c}
\hline
\hline
 Channel                   & Particle  & Exp. & \begin{tabular}[c]{@{}l@{}}Th.\\ $\chi=1$\\ $M_c^0=1.352$\\ $M_c=1.872$\end{tabular} 
 & \begin{tabular}[c]{@{}l@{}}Th.\\ $\chi=0.6$\\ $M_c^0=1.359$\\ $M_c=1.916$\end{tabular} \\ \hline
                S   & $\chi_{c0} (1P)$  & 3.415    & 3.498  & 3.448           \\
\multirow{2}{*}{PS} & $\eta_c(1S)$      & 2.984    & 3.079  & 3.022             \\
                    & $\eta_c(2S)$      & 3.638    & 3.624  & 3.600             \\
\multirow{2}{*}{V}  & $J/\Psi (1S)$     & 3.097    & 3.120  & 3.104            \\
                    & $\Psi(2S)$        & 3.686    & 3.650  & 3.650            \\
AV$_1$              & $h_c (1P)$        & 3.525    & 3.518  & 3.500            \\
AV$_2$              & $\chi_{c1} (1P)$  & 3.511    & 3.519  & 3.499         \\
T                   & $\chi_{c2} (1P)$  & 3.556    & 3.519  & 3.544    \\    
 \hline
\hline
\end{tabular}
\end{center}
\end{table}

To better understand the impact of the mixing effect, it is useful to summarize the expectation values for $\left \langle \mathbf{L\cdot S} \right \rangle$,  $\left \langle \mathbf{S_1\cdot S_2} \right \rangle$ and $\left\langle S_{12}\right\rangle$ in Tab.~\ref{tab_LSSST}. We take the mass splitting between PS and V channels, where only spin-spin interaction are operative, as an example. For simplicity, we will use the Cornell potential in Eq.~(\ref{eq_VCornellr}) to make the argument. According to Eq.~(\ref{eq_VLSSST}), the spin-spin interaction is ${V^{SS}}(r)\sim \left \langle \mathbf{S_1\cdot S_2} \right \rangle \Delta V^{vec}(r)= \left \langle \mathbf{S_1\cdot S_2} \right \rangle[\frac{16\pi\alpha_s}{3}\delta(r)+(1-\chi)\frac{2\sigma}{r} ]$. Since the quantity in the bracket is positive,  the negative (positive) value for $\left \langle \mathbf{S_1\cdot S_2} \right \rangle$ gives a more attractive (repulsive) interactions, which is the origin of the $J/\Psi$-$\eta_c$ (\ie, V-PS) splitting. The mixing with $\chi<1$
obviously enhances this effect. Similar arguments can be made for the spin-orbit, $V^{LS}(r)\sim \left \langle \mathbf{L\cdot S} \right \rangle \left [\frac{4\alpha_s}{r^3}+(3-4\chi)\frac{\sigma}{r}\right ]$, and tensor, $V^{LS}(r)\sim \left \langle S_{12} \right \rangle \left [\frac{4\alpha_s}{r^3}+(1-\chi)\frac{\sigma}{r}\right ]$, interactions; \ie, the strength of $V^{LS}$, $V^{SS}$ and $V^{T}$ are all enhanced by introducing a vector component in the confining potential, thereby improving the splittings in comparison to experiment.

\begin{table}[!t]
\tabcolsep 0pt \caption{Couplings of $LS$, $SS$ and $T$ interactions for different quarkonium channels.}
\setlength{\tabcolsep}{3pt}
\begin{center}
\def\temptablewidth{0.8\textwidth}
\begin{tabular}{c c c c c c}
\hline
\hline
  Channel        & $\left \langle \mathbf{L\cdot S} \right \rangle$      & $\left \langle \mathbf{S_1\cdot S_2} \right \rangle$   & $\left\langle S_{12}\right\rangle$       \\
   \hline
   S        & $-2$      & $1/4$  & $-4$ \\

  PS       & $0$       & $-3/4$ & $0$  \\

   V        & $0$     &  $1/4$ & $0$  \\

   AV$_1$         & $0$   & $-3/4$ & $0$    \\
   
   AV$_2$         & $-1$   & $1/4$ & $2$    \\
   
   T        & $1$   & $1/4$ & $-2/5$    \\
   \hline
\hline
\end{tabular}
\end{center}
\label{tab_LSSST}
\end{table}

\begin{figure}[htbp]
\begin{minipage}[!b]{1.0\linewidth}
\centering
\includegraphics[width=1.0\textwidth]{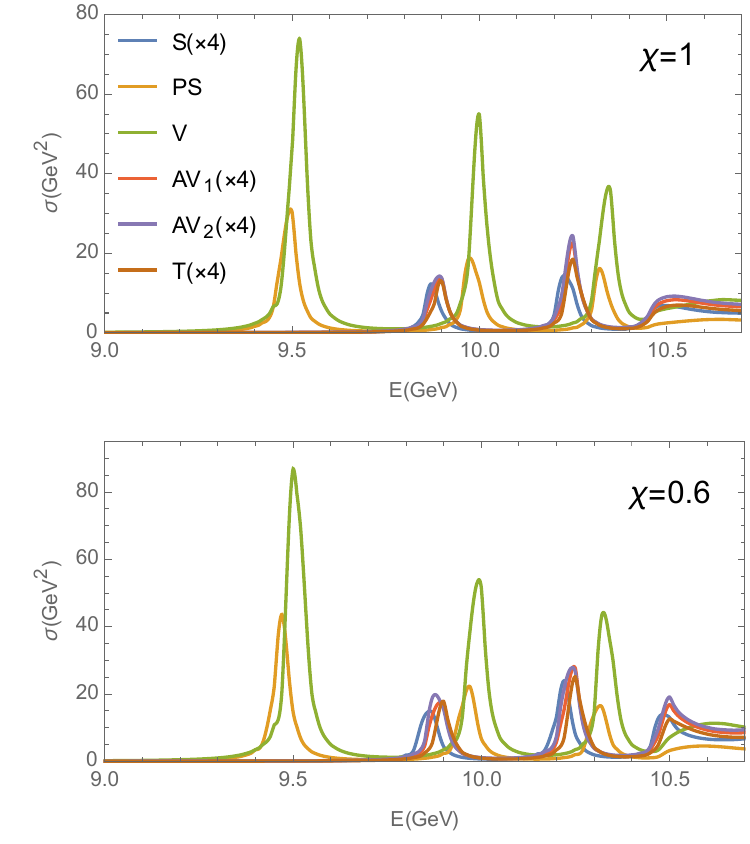}
\end{minipage}
\caption{The vacuum bottomonium spectral functions (interacting parts) in S, PS, V, AV and T channels with mixing coefficient $\chi=1$ (upper panel) and $\chi=0.6$ (lower panel). The spectral functions in S, AV and T channels are multiplied by a factor of 4 for better visibility.} 
\label{fig_specbb}
\end{figure}
We have carried out a similar analysis for bottomonium spectral functions. In Fig.~\ref{fig_specbb} bound-state spectral functions between $\chi=1$ and 0.6 are compared. We identify the peaks in each channel as follows (not all of which have an experimental counterpart (yet)): S: $\chi_{b0} (1P)$ and $\chi_{b0} (2P)$; PS: $\eta_b (1S)$, $\eta_b (2S)$ and $\eta_b (3S)$; V: $\Upsilon (1S)$, $\Upsilon (2S)$ and $\Upsilon (3S)$; AV$_1$: $h_b (1P)$, $h_b (2P)$ and $h_b (3P)$;  AV$_2$: $\chi_{b1} (1P)$, $\chi_{b1} (2P)$ and $\chi_{b1} (3P)$; T: $\chi_{b2} (1P)$, $\chi_{b2} (2P)$ and $\chi_{b2} (3P)$. The comparison between the experimental values and the masses extracted from spectral functions is compiled in Tab.~\ref{tab_bmass}. Also here an improvement in the mass splittings is found by introducing the vector confining potential, but it is not as significant as the charmonium sector, primarily due to the larger $1/M_b$ suppression for the spin-induced forces.
\begin{table}[!t]
\tabcolsep 0pt \caption{Experimental and theoretical bottomonium spectroscopy with the mixing coefficient $\chi=1$ and $\chi=0.6$ (in GeV). $M_b^0$ and $M_b$ are the bare and constituent masses for bottom quark, respectively.}
\setlength{\tabcolsep}{6pt}
\begin{center}
\def\temptablewidth{0.8\textwidth}
\begin{tabular}{c c c c c c}
\hline
\hline
 Channel  & Particle  & Exp. & \begin{tabular}[c]
{@{}l@{}}Th.\\ 
$\chi=1$\\ 
$M_b^0=4.681$\\ 
$M_b=5.247$
\end{tabular} & 
\begin{tabular}[c]
{@{}l@{}}Th.\\ 
$\chi=0.6$\\ $M_b^0=4.681$\\ $M_b=5.266$\end{tabular} \\ \hline
\multirow{2}{*}     {S}        & $\chi_{b0} (1P)$   & 9.859    & 9.871       & 9.864          \\
                               & $\chi_{b0} (2P)$   & 10.233   & 10.227      & 10.220         \\
                     PS        &$\eta_b(1S)$        & 9.399    & 9.496       & 9.470          \\
 \multirow{3}{*}    {V}        & $\Upsilon (1S)$    & 9.460    & 9.520       & 9.500           \\
                               & $\Upsilon (2S)$    & 10.023   & 9.999       & 9.994            \\
                               & $\Upsilon (3S)$    & 10.355   & 10.345      & 10.324           \\
                   AV$_1$      & $h_b (1P)$         & 9.899    & 9.896       & 9.893            \\
 \multirow{3}{*}  {AV$_2$}     & $\chi_{b1} (1P)$   & 9.893    & 9.894       & 9.877            \\
                               & $\chi_{b1} (2P)$   & 10.255   & 10.248      & 10.243           \\
                               & $\chi_{b1} (3P)$   & 10.513   & 10.520      & 10.500           \\
\multirow{2}{*}{T}             & $\chi_{b2} (1P)$   & 9.912    & 9.897       & 9.899            \\
                               & $\chi_{b2} (2P)$   & 10.269   & 10.249      & 10.249          \\  
\hline
\hline
\end{tabular}
\end{center}
\label{tab_bmass}
\end{table}

Finally, we have evaluated the spin-induced interactions in the heavy-light sector, which is the key ingredient to calculating the heavy-quark transport coefficients discussed in Sec.~\ref{sec_transport}. 
Specifically, in the $S$-wave color-singlet $D$-meson channel, the mass splitting between the pseudoscalar $D$-meson and the vector $D^*$-meson improves from 30\,MeV for $\chi$=1 to 120\,MeV for $\chi$=0.6. 

\section{In-Medium Potential and Selfconsistent QGP}
\label{sec_qgp}
In this section, we briefly introduce (and carry out) the framework for determining the medium modifications to the potential and its application to the EoS and spectral functions of the QGP within a selfconsistent quantum many-body approach~\cite{Liu:2017qah}. 
In a nutshell the procedure consists of 2 selfconsistency loops as follows.  First, the in-medium potential will be constrained through calculating the HQ free energies from the $T$-matrix and fitting it to pertinent lQCD data. The key fit parameters in this step are the screening masses, $m_d$ and $m_s$ of the color-Coulomb and string interactions. The in-medium potentials are then applied in a selfconsistent 2-PI scheme to compute the EoS of the QGP and fit those results to pertinent lQCD data as well. 
The main parameters in this step are the in-medium light-quark and gluon masses, but the EoS is computed including the full off-shell properties  of the one-body spectral functions and two-body scattering amplitudes. Since the parton selfenergies are computed from their $T$-matrices, this forms a selfconsistency problem which is solved by numerical iteration. However, the calculation of the HQ free energy also requires the spectral functions (selfenergies) of the heavy quarks, calculated from heavy-light $T$-matrices closed off with thermal parton spectral functions. Thus, after constraining the light sector with the EoS, the in-medium HQ spectral functions are re-calculated and inserted into the computation of the HQ free energies. Refitting the screening masses to the lQCD data, a refinement of the in-medium two-body potential is obtained which is then reprocessed in a new fit to the EoS. This constitutes the second (``outer") iteration loop which is also iterated numerically.   

In the remainder of this section, we first introduce the the main equations to compute the HQ free energy (Sec.~\ref{ssec_FQQ}) and the EoS (Sec.~\ref{ssec_eos}), and then discuss the numerical results with the updated in-medium potential (Sec.~\ref{ssec_self-con}).

\subsection{Static HQ Free Energy}
\label{ssec_FQQ}
Our starting point is an ansatz for the medium modifications of the potential; following previous studies~\cite{Liu:2017qah} we employ
\begin{eqnarray}
{\widetilde{V}}_{\mathcal{C}}(r)&=&-\frac{4}{3} \alpha_{s} \frac{e^{-m_{d} r}}{r}-\frac{4}{3} \alpha_{s}m_{d}\nonumber \\
{\widetilde{V}}_{\mathcal{S}}(r)&=&-\frac{\sigma e^{-m_{s} r-\left(c_{b} m_{s} r\right)^{2}}}{m_{s}}+\frac{\sigma }{m_{s}} \ , 
\label{eq_Vmedium}
\end{eqnarray}
where $m_{d}$ and $m_{s}$ are the respective Debye screening masses for Coulomb and confining potentials, related by $m_s=\left(c_{s} m_{d}^2 \sigma/\alpha_s\right)^{1/4}$~\cite{Liu:2017qah}. The quadratic term in the exponential, $-\left(c_{b} m_{s} r\right)^{2}$, accelerates the suppression of the long-range part of the confining potential to simulate string breaking. In the limit of vanishing screening masses the vacuum potential of Eq.~(\ref{eq_VCornellr}) is recovered.

The HQ free energy, $F_{Q\bar{Q}}(r,T)$, is defined as the difference between the free energies of the QGP without and with a static quark and antiquark (not counting their infinite masses) separated by a distance $r$ (see, \eg, Ref.~\cite{Beraudo:2007ky}),
\begin{equation}
F_{Q \bar{Q}}(r, T)=-\frac{1}{\beta} \ln \left[G_{Q \bar{Q}}^{>}(-i \beta, r)\right],
\end{equation}
where $G_{Q \bar{Q}}^{>}(-i \tau, r)$ is the Euclidean time Green function and $\beta=1/T$ the inverse temperature. In the vacuum, this simply corresponds to the potential between $Q$ and $\bar Q$, cf. Sec.~\ref{sec_Vvac}. In medium, the free energy and the potential are no longer identical to each other due to the presence of entropy contributions resulting from medium effects encoded in the HQ selfenergies (which we calculate from the in-medium heavy-light $T$-matrix) and the potential. In Ref.~\cite{Liu:2017qah} a compact form of the free energy has been derived as
\beq
\begin{aligned}
F_{Q \bar{Q}}(r, T)=& -\frac{1}{\beta} \ln \left[-\int_{-\infty}^{\infty} \frac{dE}{\pi} e^{-\beta E} \right.
\\
&\left.\times \operatorname{Im}\left[\frac{1}{E+i \epsilon-\widetilde{V}(r)-\Sigma_{Q \bar{Q}}(E+i \epsilon)}\right]\right],
\label{eq_free energy}
\end{aligned}
\eeq
with the color-singlet potential $\widetilde{V}(r)$ (color-flavor indices are suppressed for simplicity) from Eq.~(\ref{eq_Vmedium}). The relationship between the two-body selfenergy, $\Sigma_{Q \bar{Q}}(z)$, and the two-body propagator, $G_{Q \bar{Q}}^{0}(z)$, is~\cite{Liu:2017qah} 
\begin{equation}
\left[G_{Q \bar{Q}}^{0}(z)\right]^{-1}=z-2 \Delta M_{Q}-\Sigma_{Q \bar{Q}}(z) \ ,
\label{eq_2bodyselfenergy}
\end{equation}
with a Fock mass term for each quark, $ \Delta M_{Q}=\widetilde{V}(\infty)/2$. In the static limit, $G_{Q \bar{Q}}^{0}(z)$ reduces to 
\begin{equation}
G_{Q \bar{Q}}^{0}(z)=\int_{-\infty}^{\infty} d \omega_{1} d \omega_{2} \frac{\rho_{Q}\left(\omega_{1}\right) \rho_{\bar{Q}}\left(\omega_{2}\right)}{z-\omega_{1}-\omega_{2}} \ ,
\label{eq_G00static}
\end{equation}
where $\rho_{Q/\bar{Q}}(\omega)=-\frac{1}{\pi} \operatorname{Im} G_{Q/\bar{Q}}(\omega+i \epsilon)$ are the single-particle spectral functions with propagators $G_{Q/\bar{Q}}(z)=\frac{1}{z-M_{Q/\bar{Q}}-\Sigma_{Q/\bar{Q}}(z)}$ in the static limit. Then the single-particle selfenergy, $\Sigma_{Q}(z)$, can be solved selfconsistently by combining the $T$-matrix and the selfenergy equations. By taking the heavy-light $T$-matrix from Eq.~(\ref{eq_3DTM}) in the ``half-static" limit, where the $\mathbf{p_1}$ dependence is suppressed due to the infinite static-quark mass, Eq.~(\ref{eq_selfenergy}) takes the form
\begin{equation}
\begin{aligned}
\Sigma_{Q}(z)=& \int \frac{d^{3} \mathbf{p}_{2}}{(2 \pi)^{3}} \int_{-\infty}^{\infty} d \omega_{2} \frac{d E}{\pi} \frac{-1}{z+\omega_{2}-E} \frac{1}{d_{Q}} \sum_{a, j} d_{s}^{Q j} d_{a}^{Q j} \\
& \times T_{Q j}^{a}\left(E, \mathbf{p}_{2} \mid \mathbf{p}_{2}\right) \rho_{j}\left(\omega_{2}, \mathbf{p}_{2}\right) n_{j}(\omega_{2}) \ .
\end{aligned}
\end{equation}
The CM transformation in Eq.~(\ref{eq_trans}) reduces to 
\begin{equation}
E_{\mathrm{cm}}=\omega_{1}+\omega_{2}, \quad p_{\mathrm{cm}}=p_{2}, \quad \cos \left(\theta_{\mathrm{cm}}\right)=\cos (\theta),
\end{equation}
with $\omega_1+\omega_2\gg|\mathbf{p}_{1}+\mathbf{p}_{2}|$. The resulting single-particle selfenergy, $\Sigma_{Q}(z)$, is inserted into Eq.~(\ref{eq_G00static}) to obtain the $Q\bar Q$ propagator, and Eq.~(\ref{eq_2bodyselfenergy}) yields the two-body selfenergy, $\Sigma_{Q \bar{Q}}(z)$.

Interference effects lead to a suppression of the imaginary part of the two-body selfenergy (relative to the sum of the single-particle absorptive parts), which is sometimes referred to as ``imaginary part of the potential" (it is, in fact,  an $r$-dependent suppression of the imaginary part)~\cite{Laine:2006ns}. In the $T$-matrix formalism this amounts to 3-body diagrams which are rather challenging to compute explicitly~\cite{Liu:2017qah}. Instead, the interference effects are implemented through an $r$-dependent suppression factor~\cite{Liu:2017qah} with a functional form guided by perturbative results~\cite{Akamatsu:2020ypb} using a factorized ansatz, $\Sigma_{Q \bar{Q}}(z,r)=\Sigma_{Q \bar{Q}}(z)\phi(r)$, where the function $\phi(r)$ is part of the constraints from the lQCD data for static HQ free energies at each temperature. The interference effect is mostly relevant for deeply bound heavy quarkonia, where, in the color singlet channel, the imaginary part should vanish in the limit of $r\to0$ (corresponding to a color-neutral object). This is a central ingredient to quantum transport approaches (see Ref.~\cite{Akamatsu:2020ypb} for a recent review), but it also plays a significant role in the quantitative description of the quarkonium spectral functions computed within the $T$-matrix approach, especially when fitting lQCD data for euclidean quarkonium correlators~\cite{Liu:2017qah}.

\subsection{Equation of State}
\label{ssec_eos}
The equation of state of a many-body system is encoded in the pressure, $P(T,\mu)$, as a function of temperature and chemical potential. The EoS is driven by the dominant degrees of freedom in the medium, and is therefore sensitive to their spectral properties, including their masses. For a homogeneous grand canonical ensemble, the relationship between the EoS and the grand potential per unit volume is given by $\Omega=-P$. We adopt the Luttinger-Ward-Baym (LWB) formalism which provides a diagrammatic and thermodynamically consistent quantum approach that allows to incorporate the off-shell dynamics of the one- and two-body correlation functions. Quantum effects are expected to become particularly important for a strongly coupled system with large scattering rates (widths)~\cite{Luttinger:1960ua,Baym:1961zz,Baym:1962sx}. One has
\begin{equation}
\Omega=\mp \frac{-1}{\beta} \sum_{n} \operatorname{Tr}\left\{\ln \left(-G^{-1}\right)+\left[\left(G^{0}\right)^{-1}-G^{-1}\right] G\right\} \pm \Phi,
\label{eq_LWB}
\end{equation}
where ``$\operatorname{Tr}$" denotes the trace over spin, color, flavor and 3-momentum, $\sum_{n}$ the Matsubara frequency sum, and the $G^0$ and $G$ are the free and fully dressed single-particle Green's function. The two-body interaction contribution is encoded in the  Luttinger-Ward functional (LWF), $\Phi=\sum_{v=1}^{\infty} \Phi_{v}$, where
\begin{equation}
\Phi_{v}=\frac{-1}{\beta} \sum_{n} \operatorname{Tr}\left\{\frac{1}{2 v}\left(\frac{-1}{\beta}\right)^{v}\left[(-\beta)^{v} \Sigma_{v}(G)\right] G\right\}
\label{eq_LWF1}
\end{equation}
with
\begin{equation}
\Sigma_{v}(G)=\int d \tilde{p}\left[V G_{(2)}^{0} V G_{(2)}^{0} \cdots V\right] G\ ,
\label{eq_LWF2}
\end{equation}
using the notation $\int d \tilde{p} \equiv-\beta^{-1} \sum_{n} \int d^{3} \mathbf{p} /(2 \pi)^{3}$ with $\tilde{p} \equiv\left(i \omega_{n}, \mathbf{p}\right)$. The $\phi_\nu$ correspond to the ``skeleton diagrams" of the $\nu^{\rm th}$ order in the potential expansion. To account for possible bound-states formation and their contribution to the pressure, one has to resum the skeleton series. For non-separable interactions this has been achieved through a matrix-logarithm resummation technique in Refs.~\cite{Liu:2017qah,Liu:2016nwj,Liu:2016ysz}, resulting in a structure similar to the $T$-matrix resummation in Eq.~(\ref{eq_selfenergy}):
\begin{equation}
\begin{aligned}
\Omega=& \sum_{j} \mp d_{j} \int d \tilde{p}\left\{\ln \left(-G_{j}(\tilde{p})^{-1}\right)\right.\\
&\left.+\left[\Sigma_{j}(\tilde{p})-\frac{1}{2} \log \Sigma_{j}(\tilde{p})\right] G_{j}(\tilde{p})\right\}
\end{aligned}
\label{eq_grand}
\end{equation}
with
\begin{equation}
\begin{aligned}
\log \Sigma_{i}\left(z, \mathbf{p}_{1}\right)&= \int \frac{d^{3} \mathbf{p}_{2}}{(2 \pi)^{3}} \int_{-\infty}^{\infty} d \omega_{2} \frac{d E}{\pi} \frac{-1}{z+\omega_{2}-E} 
\\
&  \times \frac{1}{d_{i}} \sum_{a, j} d_{s}^{i j} d_{a}^{i j} \operatorname{Im}\left[\log T_{i j}^{a}\left(E, \mathbf{p}_{1}, \mathbf{p}_{2} \mid \mathbf{p}_{1}, \mathbf{p}_{2}\right)\right] \\
& \times \rho_{j}\left(\omega_{2}, \mathbf{p}_{2}\right)\left[n_{j}\left(\omega_{2}\right) \mp n_{i j}(E)\right].
\end{aligned}
\label{eq_logselfenergy}
\end{equation}
The transformation of the $T$-matrices in Eq.~(\ref{eq_logselfenergy}) between the thermal and CM frame is given by Eq.~(\ref{eq_trans}). The grand potential can then be obtained after carrying out the sum over Matsubara frequencies in Eq.~(\ref{eq_grand}). 

\subsection{Selfconsistent in-Medium Results}
\label{ssec_self-con}
We now turn to the selfconsistent in-medium results at four temperatures, $T=0.194$~GeV, 0.258~GeV, 0.320~GeV and 0.400~GeV, constrained by the lQCD data for static HQ free energies  (Sec.~\ref{sssec_res-FQQ}) and QGP EoS (Sec.~\ref{sssec_res-eos}). All in-medium calculations are carried out with the mixing coefficient for $\chi=0.6$ and 1 in this study, but we do not yet incorporate the spin-dependent corrections. In particular in the light sector, \ie, for the QGP EoS, their effect can be rather significant and deserves a separate study (some compensatory effects are expected due to both attractive and repulsive contributions). Nevertheless, we want to ensure that the medium within which the heavy quarks are embedded satisfies basic constraints from lQCD. 

\subsubsection{Static HQ Free Energies}
\label{sssec_res-FQQ}
Recalling Eq.~(\ref{eq_free energy}), the HQ free energy, $F_{Q\bar{Q}}(r,T)$, is a functional of the potential, $\widetilde{V}(r)$, and the two-body selfenergy, $\Sigma_{Q\bar{Q}}(E+i \epsilon)$. Note that $F_{Q\bar{Q}}(r,T)$ increases with increasing $\widetilde{V}(r)$ but with decreasing $|\Sigma_{Q}(E+i \epsilon)|$. A larger Debye screening mass, $m_d$ and/or $m_s$, suppresses $\widetilde{V}(r)$  so that the partons become more weakly coupled, which in turn lowers $F_{Q\bar{Q}}(r,T)$; at the same time, a larger $m_d$ reduces $|\Sigma_{Q}(E+i \epsilon)|$ in medium, which in turn enhances $F_{Q\bar{Q}}(r,T)$. It is this competition between $\widetilde{V}(r)$ and $|\Sigma_{Q}(E+i \epsilon)|$ that leads to a non-monotonic behavior of $F_{Q\bar{Q}}(r,T,m_d)$ with $m_d$. Since $m_d$ is directly related to the free energy at infinite distance (cf.~Sec.~\ref{ssec_FQQ}), we define a function $F_{trial}(r\rightarrow\infty,T,m_d)$ calculated from our many-body approach which we require to be equal to the lQCD value, $F_{lQCD}(r\rightarrow\infty,T)$. We typically find two solutions for $m_d$ for any fixed parameter set provided the maximum of the trial free energy lies above the lQCD value. We denote the solutions with the smaller and the larger $m_d$ as strongly coupled solution (SCS) and weakly coupled solution (WCS), respectively (in analogy to the two solutions which were found in Ref.~\cite{Liu:2017qah}). Here, we only focus on the SCS which results in transport parameters in much better agreement with heavy-ion phenomenology (\ie, a liquid-like behavior with interaction energies comparable to the parton masses, as well as HQ transport parameters) than the WCS~\cite{Liu:2017qah}.

The resulting potentials and fits to lQCD results~\cite{Bazavov:2012fk} for the HQ free energies with $c_b=1.55$ (1.3) and $c_s=0.06$ (0.01) for $\chi=0.6$ (1) at different temperatures are shown in Fig.~\ref{fig_free energy}. 
As found in earlier studies, large HQ widths lead to a substantial enhancement of the potential over the free energies; in particular, at the lowest temperature of 0.194\,GeV, the potential is close to the vacuum one, but becomes notably suppressed at higher $T$.   
\begin{figure*}[htbp]
\begin{minipage}[b]{1.0\linewidth}
\centering
\includegraphics[width=1.0\textwidth]{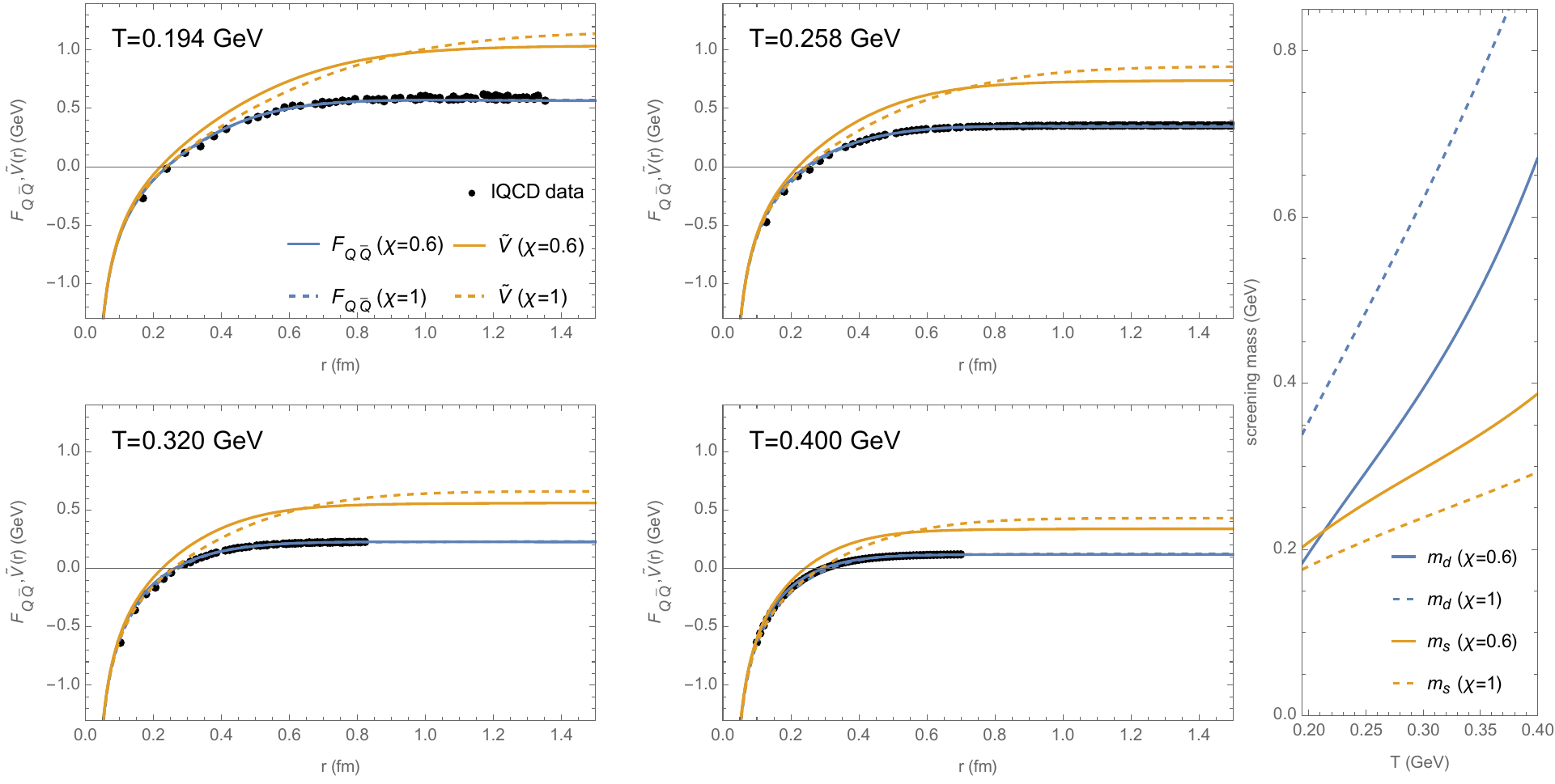}
\end{minipage}
\caption{Left and middle panels: The in-medium HQ free energies (blue) and potentials (orange) for $\chi=0.6$ (solid) and 1 (dashed) at different temperatures in comparison to lQCD data for the HQ free energies from Ref.~\cite{Bazavov:2012fk} for $N_f=2+1$ light flavors (black dots). The right panel shows the temperature dependence of the 
screening masses, $m_d$ (blue) and $m_s$ (orange), for $\chi=0.6$ (solid) and 1 (dashed) as resulting from our fit. The $\chi=1$ results are taken from Ref.~\cite{Liu:2017qah}.} 
\label{fig_free energy}
\end{figure*}
Consequently, the screening masses for Coulomb ($m_d$) and confining ($m_s$) potentials, shown in the right panel of Fig.~\ref{fig_free energy}, have rather small values at low $T$, with the string interaction exhibiting a weaker screening with increasing $T$. This implies that remnants of the confining force survive in the QGP well above the critical region. The potential with mixed confining interaction ($\chi=0.6$) is enhanced by the extra relativistic corrections (cf.~Eq.~(\ref{eq_R correction})), requiring a stronger screening to fit the lQCD free-energy data. Therefore, the screening masses for confining potential for $\chi=0.6$ are larger than those for $\chi=1$, see the right panel of Fig.~\ref{fig_free energy}. However, note that the $\chi=0.6$ solution generates a stronger force at relatively small distances, a feature that will figure importantly in the QGP structure and HQ transport properties.

\subsubsection{Equation of State}
\label{sssec_res-eos}
In  Fig.~\ref{fig_eos} we display the pressure together with the fitted light-parton masses for $\chi=0.6$ and 1, which allow for a good reproduction of the lQCD data. However,  while the parton masses are effective in achieving a quantitative agreement with the lQCD results, the underlying quark and gluon spectral functions for $\chi=0.6$ and 1 both feature large selfenergies, especially imaginary parts which, at low momentum and temperatures,  are comparable to, or even larger, than the parton masses, cf.~the spectral function widths in Fig.~\ref{fig_rho-qgc} (first and second rows). 
The large scattering rates are mostly driven by dynamical resonance formation in the underlying  $T$-matrices (which in turn are generated by resumming the strong potential). These resonances contribute through the resummed LWF functional $\Phi\sim1/2 \textup{log}\Sigma G$ introduced in Sec.~\ref{ssec_eos}, whose contribution for  $\chi=0.6$ and 1 is displayed in Fig.~\ref{fig_eos}. The increasing proportion of LWF contribution with decreasing temperature indicates the onset of a change in the degrees of freedom. Specifically, the LWF parts make up more than 70(50)\% of the pressure at $T=0.194$\,GeV for $\chi=0.6(1)$. 
While the spectral functions for $\chi=0.6$ generally share the main features with those for $\chi=1$ at low momenta, a notable quantitative difference is that the widths for $\chi=0.6$ do not fall off with momentum as much as those for $\chi=1$. In the former case, this is a consequence of the 3-momentum dependence of the confining interaction whose vector component, through relativistic effects, generates additional interaction strength and thus larger scattering rates at larger momenta relative to the  $\chi=1$ case.

\begin{figure}[htbp]
\begin{minipage}[b]{1.0\linewidth}
\centering
\includegraphics[width=1.0\textwidth]{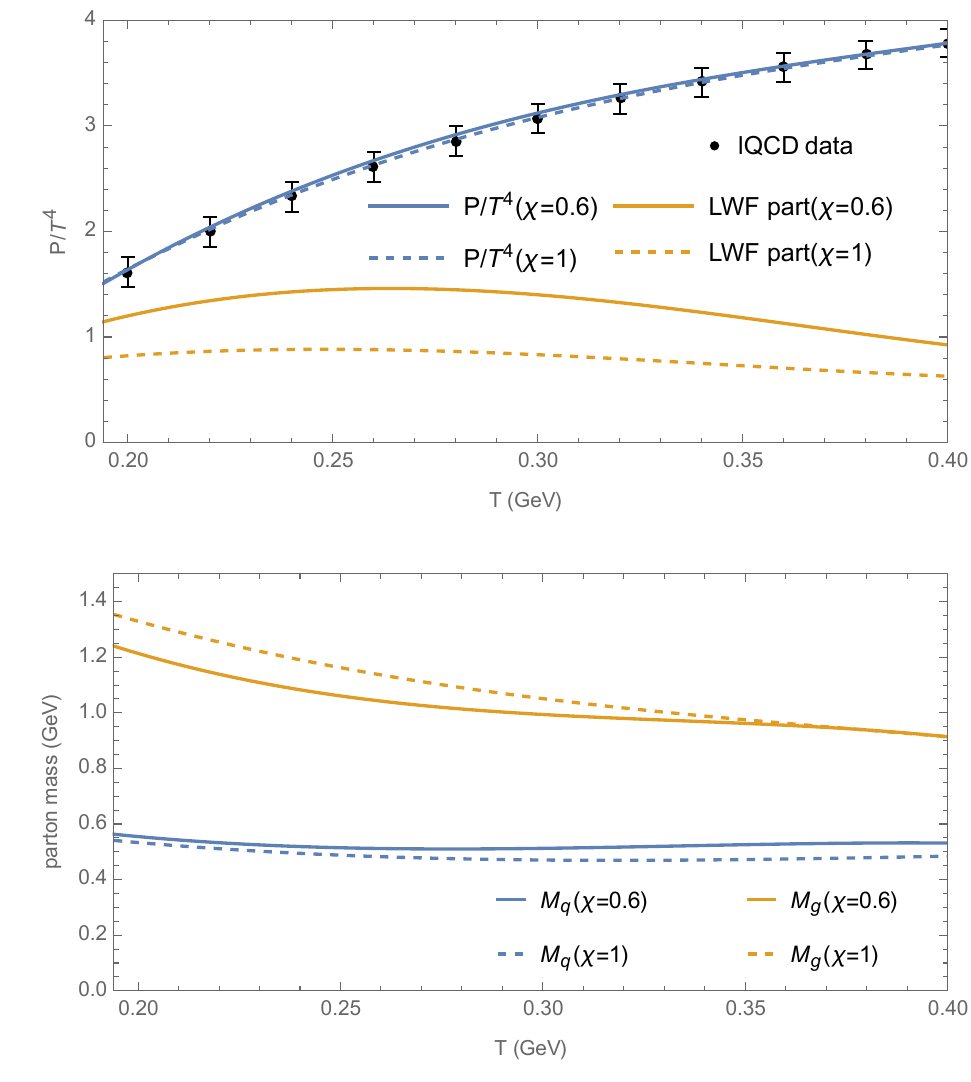}
\end{minipage}
\caption{The pressure (normalized by $T^4$) in comparison to the lQCD data (black dots) from Ref.~\cite{HotQCD:2014kol} (upper panel), and the in-medium light-quark and gluon masses as a function of temperature (lower panel). The $\chi=1$ results are taken from Ref.~\cite{Liu:2017qah}.} 
\label{fig_eos}
\end{figure}

\begin{figure*}[htbp]
\begin{minipage}[b]{1.0\linewidth}
\centering
\includegraphics[width=1.0\textwidth]{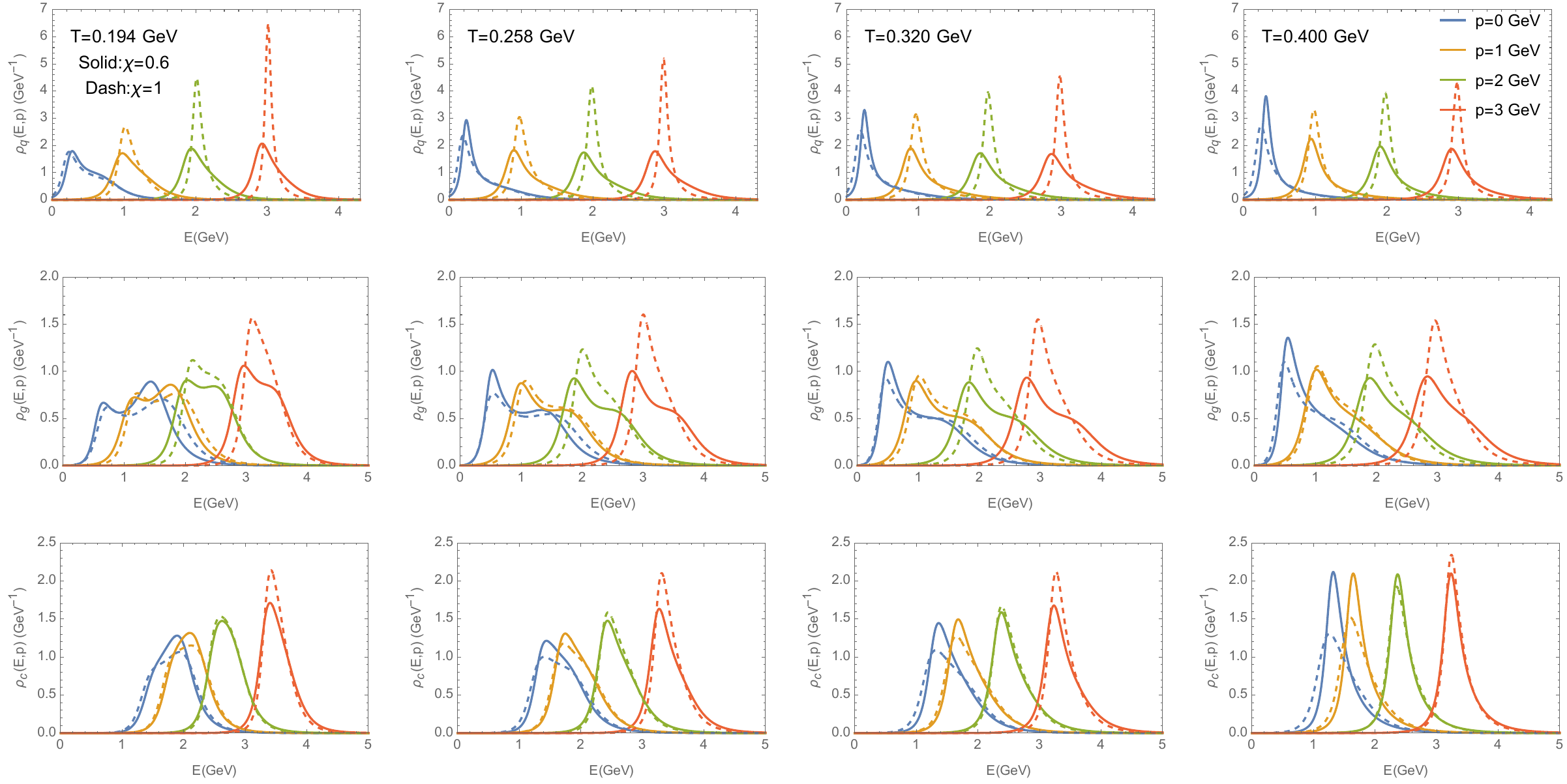}
\end{minipage}
\caption{Single-parton spectral functions for light quarks (first row), gluons (second row) and charm quarks (third row) with $\chi=0.6$ (solid) and 1 (dashed) as a function of energy for various 3-momenta in each panel. From left to right, the four columns correspond to temperatures of $T=194$, 258, 320 and 400\,MeV, respectively. The $\chi=1$ results are taken from Ref.~\cite{Liu:2017qah}.} 
\label{fig_rho-qgc}
\end{figure*}
  
\section{Charm-Quark Transport Coefficients}
\label{sec_transport}
With the parameters of the interaction potential and parton masses fixed with the aid of lQCD data, we can now investigate the effect of the mixed potential on charm-quark transport properties in the QGP. As elaborated in Ref.~\cite{Liu:2018syc} it is important to account for the off-shell properties of both charm quarks and thermal partons in the evaluation of the transport coefficient, especially due to the formation of near-threshold bound states which only provide limited phase for quasiparticle (on-shell) scattering. This point is further corroborated upon inspecting the equilibrium spectral functions, $\rho_{q,g,c}$, of the partons displayed in Fig.~\ref{fig_rho-qgc}, exhibiting large widths of $\sim$0.6\,GeV or so at low momentum.  As already mentioned in Sec.~\ref{sssec_res-eos}, the main difference between $\chi=0.6$ and 1 is that the widths for $\chi=0.6$ do not fall off with momentum as much as those for $\chi=1$.
This feature persists in the heavy-light scattering amplitudes, which are the main ingredient to the charm-quark transport coefficients discussed below, see Fig.~\ref{fig_Thl}. At $T=0.194$\,GeV, the peak value of the imaginary part of the $S$-wave color-singlet heavy-light scattering amplitude for $\chi=0.6$ still shows a rather marked decrease with increasing center-of-mass mass momentum of the colliding  partons, but it is significantly weaker than for $\chi=1$ with a purely scalar confining potential; \eg, the peak reduction from the $p_{cm}=0$ to $p_{cm}=0.5$\,GeV case is almost a factor 3 for the latter but only $\sim$1.5 for $\chi=0.6$. This trend continues to higher temperatures; at $T=0.400$\,GeV,  the peak reduction from  $p_{cm}=0$ to $p_{cm}=0.6$\,GeV is essentially absent for $\chi=0.6$, while it is still a factor of 1.6 for the purely scalar confining potential. The $T$-matrix amplitudes for $\chi=0.6$ are smaller than that for $\chi=1$ at low momenta due to its stronger screening in confining potential (recall its larger Debye screening masses in the right panel of Fig.~\ref{fig_free energy}); however, they exceed the ones for $\chi=1$ for 
$p_{cm}\gtsim$0.5\,GeV due to the harder 3-momentum dependence of confining potential through relativistic effects. The $T$-matrices for other partial waves and color channels share similar features, and thus we do not reproduce therm here.
\begin{figure*}[htbp]
\begin{minipage}[b]{1.0\linewidth}
\centering
\includegraphics[width=1.0\textwidth]{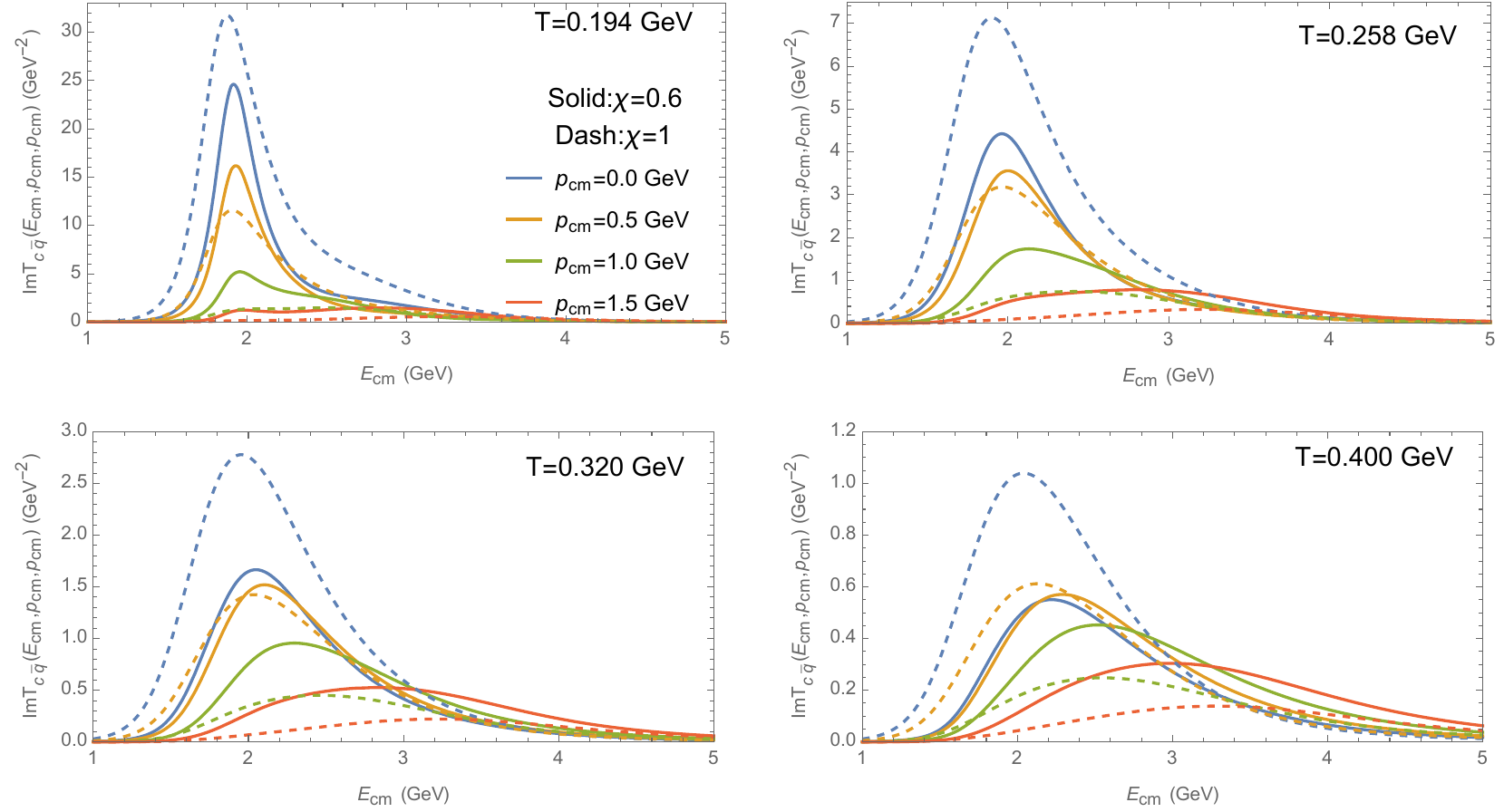}
\end{minipage}
\caption{The imaginary part of the $S$-wave charm-light $T$-matrices in the color-singlet channel at different temperatures. The $T$-matrix is displayed for four different values of the CM momentum ($p_{cm}$) in each panel. The $\chi=1$ results are taken from Ref.~\cite{Liu:2017qah}.} 
\label{fig_Thl}
\end{figure*}

Turning now to the HQ transport coefficients in the QGP, we adopt their standard definition through a Fokker-Planck equation where they amount to the first and second momentum of the momentum transfer of the heavy-light scattering amplitude squared, integrated over the thermal-parton distributions (one could also employ a Kubo-type formula via the zero-mode contribution to the charmonium spectral function in the vector channel, see, \eg, Ref.~\cite{Riek:2010py}). At this level, off-shell effects can be readily implemented by an additional energy convolution over the light-parton spectral functions. However, since also charm quarks acquire widths which are not small, the inclusion of their spectral width is also warranted. This has been worked out in Ref.~\cite{Liu:2018syc} employing the Kadanoff-Baym equations, resulting in the following expression for the HQ friction coefficient (or relaxation rate): 
\begin{equation}
\begin{aligned}
A(p)=& \sum_{i} \frac{1}{2 \varepsilon_{c}(p)} \int \frac{d \omega^{\prime} d^{3} \mathbf{p}^{\prime}}{(2 \pi)^{3} 2 \varepsilon_{c}\left(p^{\prime}\right)} \frac{d \nu d^{3} \mathbf{q}}{(2 \pi)^{3} 2 \varepsilon_{i}(q)} \frac{d \nu^{\prime} d^{3} \mathbf{q}^{\prime}}{(2 \pi)^{3} 2 \varepsilon_{i}\left(q^{\prime}\right)} \\
& \times \delta^{(4)} \frac{(2 \pi)^{4}}{d_{c}} \sum_{a, l, s}|M|^{2} \rho_{c}\left(\omega^{\prime}, p^{\prime}\right) \rho_{i}(\nu, q) \rho_{i}\left(\nu^{\prime}, q^{\prime}\right) \\
& \times\left[1-n_{c}\left(\omega^{\prime}\right)\right] n_{i}(\nu)\left[1 \pm n_{i}\left(\nu^{\prime}\right)\right] (1-\frac{\mathbf{p}\cdot\mathbf{p'}}{\mathbf{p}^2}) \ .
\end{aligned}
\label{eq_A(p)}
\end{equation}
As before (recall Sec.~\ref{sec_TM}) ,  $\varepsilon_{i/c}$, $\rho_{i/c}$ and $n_{i/c}$ are the dispersion relations, spectral and thermal-distribution functions for partons $i/c$, respectively, $\delta^{(4)}$ is a short-hand notation for the energy-momentum conserving $\delta$-function in the 2$\to$2 scattering process, and $d_c=6$ the spin-color degeneracy of charm quarks. The summation $\sum_{i}$ is over all light-flavor quarks and gluons, $i=u,\bar{u},d,\bar{d},s,\bar{s},g$, where the masses for light and strange quarks are assumed to be the same. In the above expression, the quasiparticle approximation is only applied to the incoming charm quark by assigning it a sharp energy $\varepsilon_{c}(\mathbf{p})$ at momentum $\mathbf{p}$, while all other partons are treated via off-shell integrations. We expect this approximation to be reasonable for charm-quark widths that can be larger than the temperature but are still smaller than the charm-quark on-shell energy, which is in practice the case for the interactions considered here.
The heavy-light scattering matrix elements, $|M|^2$, in Eq.~(\ref{eq_A(p)}) are related to the $T$-matrix in the CM frame by
\begin{equation}
\begin{aligned}
&\sum_{a, L, s}\left|M\right|^{2}=16 \varepsilon_{c}\left(p_{\mathrm{cm}}\right) \varepsilon_{i}\left(p_{\mathrm{cm}}\right) \varepsilon_{c}\left(p_{\mathrm{cm}}^{\prime}\right) \varepsilon_{i}\left(p_{\mathrm{cm}}^{\prime}\right) d_{s}^{c i} \\
&\times \sum_{a} d_{a}^{ci}\left|4 \pi \sum_{L}(2 L+1) T_{ci}^{a, L}\left(E_{\mathrm{cm}}, p_{\mathrm{cm}}, p_{\mathrm{cm}}^{\prime}\right) P_{L}(x)\right|^{2}
\end{aligned}
\label{eq_M2}
\end{equation}
with the color and spin degeneracies of the two-body system, $d_{a,s}^{ci}$. The heavy-light $T$-matrix, 
$T_{ci}^{a, L}(E_{\mathrm{cm}},p_{\mathrm{cm}},p_{\mathrm{cm}}^{\prime})$, is calculated in the CM frame in all possible two-body color 
channels, $a$, and partial-wave channels, $L$ (expanded up to $L=8$ to ensure convergence at high momenta). The CM energy $E_{\mathrm{cm}}$, incoming CM momentum $p_{\mathrm{cm}}$, outgoing CM momentum $p'_{\mathrm{cm}}$, and scattering angle, $x=\cos \theta_{\mathrm{cm}}$, are expressed as functions of $E, \mathbf{p}, \mathbf{q}, \mathbf{p}', \mathbf{q}'$, through the transformation in Eq.~(\ref{eq_trans}). Instead of only the moduli of $p_{\mathrm{cm}}$ and $p'_{\mathrm{cm}}$, their explicit vector form is required here~\cite{Liu:2018syc}:
\begin{equation}
\begin{aligned}
&p_{\mathrm{cm} \|}=\frac{\varepsilon_{\mathrm{p}_{2}} p_{1 \|}-\varepsilon_{\mathrm{p}_{1}} p_{2 \|}}{\sqrt{s_{\mathrm{on}}}}, & \mathbf{p}_{\mathrm{cm} \perp}=\frac{\mathbf{p}_{1} p_{2 \|}-\mathbf{p}_{2} p_{1 \|}}{\left|\mathbf{p}_{1}+\mathbf{p}_{2}\right|},
\end{aligned}
\end{equation}
with $\|$ and $\perp$ indicating parallel and perpendicular to the relative velocity, respectively, and likewise for the outgoing (primed) momenta. 

\begin{figure}[htbp]
\begin{minipage}[b]{1.0\linewidth}
\centering
\includegraphics[width=1.0\textwidth]{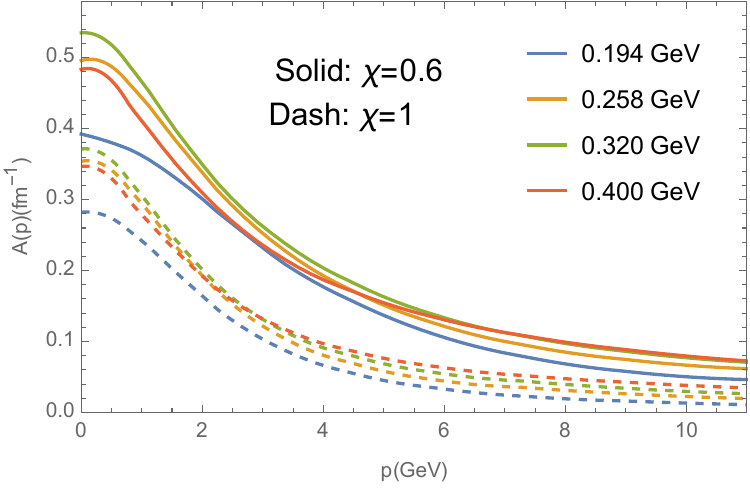}
\end{minipage}
\caption{The charm-quark friction coefficient at different temperatures for $\chi=0.6$ (solid) and 1 (dashed). The $\chi=1$ results are taken from Ref.~\cite{Liu:2018syc}.} 
\label{fig_Ap}
\end{figure}
In Fig.~\ref{fig_Ap} we plot our results for the friction coefficient $A(p)$ with the mixed confining potential ($\chi=0.6$) in comparison to the results with a purely scalar confining potential ($\chi=1$). We stipulate that both calculations are carried out for a thermal QGP medium which satisfies the constraints from the EoS and HQ free energy. With the vector component in the confining potential, the low-momentum values of the relaxation rate are enhanced by several tens of percent, but the more significant effect is the increase at higher momenta, for the same reasons as discussed above in the context of the single-parton spectral functions and their scattering amplitudes. For example, for a charm-quark momentum of 4\,GeV, the enhancement is about a factor 2.6, while at momenta of 10\,GeV it reaches an even larger factor of $\sim$3.5 at the lowest temperature. However, at the latter momentum, radiative contributions are expected to be large. At first sight it might be surprising that the enhancement due to the vector component in the confining potential also transpires at low momenta although the pertinent $T$-matrix amplitudes are smaller than those with purely scalar confining potential at low CM momenta (cf. Fig.~\ref{fig_Thl}). To some extent this can be understood due to the fact that even at vanishing HQ momentum the thermal motion of the surrounding medium partons creates a finite momentum in the CMS, but there is also a non-trivial interference effect in the expression (\ref{eq_M2}) that plays a role. 

To scrutinize different contributions, we take the charm-light contribution ($c\bar{q}$) for $A(p=0)$ at $T=194$\,MeV as an example and collect in Tab.~\ref{tab_rates} partial-wave components of the collision rate (obtained by replacing $(1-\frac{\mathbf{p}\cdot\mathbf{p'}}{\mathbf{p}^2})$ by 1 in Eq.~(\ref{eq_A(p)})) and relaxation rate up to angular momenta of 2 (note that for the collision rate the interference contributions should vanish, whereby the small numerical values quoted in the table, which are of the order of 1-2 permille of the total, are an indication of our numerical accuracy) . We denote by ``$LL'$" the terms $\sim T^{L}(T^{L'})^*=\textup{Re}T^{L}\textup{Re}T^{L'}+\textup{Im}T^{L}\textup{Im}T^{L'}+i(-\textup{Re}T^{L}\textup{Im}T^{L'}+\textup{Im}T^{L}\textup{Re}T^{L'})$ in Eq.~(\ref{eq_M2}), where the imaginary part vanishes by definition since $T^{L}(T^{L'})^*+(T^{L})^*T^{L'} = 2(\textup{Re}T^{L}\textup{Re}T^{L'}+\textup{Im}T^{L}\textup{Im}T^{L'})$. In accord with the $T$-matrix behavior at low momenta in Fig.~\ref{fig_Thl}, the collision rate for $\chi=1$ is larger than that for $\chi=0.6$ for each partial-wave component, leading to a larger total collision rate at low momentum. 
The situation is more involved for the relaxation rate: the diagonal partial-wave components ($L=L'$) for $\chi=1$ are smaller than those for $\chi=0.6$, and one also notices the relatively more important role of the higher partial waves compared to the collision rate (which is dominated by the $S$-wave contribution). In addition, the presence of the ${\mathbf{p}\cdot\mathbf{p'}}$ term, together with the Legendre polynomials, causes negative interference components ($L\neq L'$), and their absolute values are larger for $\chi=1$. Upon adding the diagonal and interference components the relaxation rate for $\chi=0.6$ becomes larger. 

The widely discussed spatial diffusion coefficient, $D_s=T/(M_c A(p=0))$, is related to the relaxation time, $\tau_c = 1/A(p=0)$, at vanishing 3-momentum of the heavy quark.
It is commonly scaled by the inverse thermal wavelength, $2\pi T$, to render a dimensionless quantity for which we display our results in Fig.~\ref{fig_Ds} as a function of temperature. The $\chi=0.6$ result shows a mild reduction relative to the $\chi=1$ one, which is again caused by the larger average momenta of the thermal partons probed by the charm quark.

\begin{table}[ht]
\tabcolsep 0pt \caption{The contributions of various partial-wave components (specified in the first row) of the $c\bar{q}$ scattering amplitude to the $c$-quark collision rate (lines 2 and 3) and relaxation rate (lines 4 and 5) for $p=0$ at $T=194$\,MeV (in units of $\operatorname{fm^{-1}}$).}
\setlength{\tabcolsep}{3pt}
\begin{center}
\def\temptablewidth{0.8\textwidth}
\begin{tabular}{c c c c c c}
\hline
\hline
 $LL'$      & $00$      & $11$      & $22$     & $01$=$10$  & $12$=$21$     \\
   \hline
 $\chi=1$   & $0.8058$  & $0.3321$  & $0.0678$ & $0.0038$   & $0.0016$\\

 $\chi=0.6$ & $0.7308$  & $0.2943$  & $0.0533$ & $0.0022$   & $0.0014$\\
   \hline
\hline
\hline
 $\chi=1$   & $0.1501$  & $0.1138$  & $0.0354$ & $-0.0521$   & $-0.0306$\\

 $\chi=0.6$ & $0.1522$  & $0.1313$  & $0.0359$ & $-0.0322$   & $-0.0305$\\
   \hline
\hline
\end{tabular}
\end{center}
\label{tab_rates}
\end{table}
\begin{figure}[htbp]
\begin{minipage}[b]{1.0\linewidth}
\centering
\includegraphics[width=1.0\textwidth]{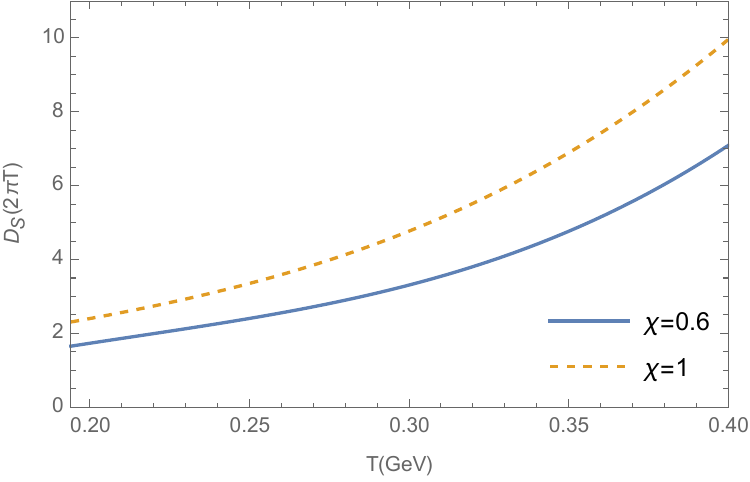}
\end{minipage}
\caption{The charm-quark spatial diffusion coefficient for $\chi=0.6$ (solid) and 1 (dashed). The $\chi=1$ results are taken from Ref.~\cite{Liu:2018syc}.} 
\label{fig_Ds}
\end{figure}

The increase in the elastic charm-quark friction coefficient, and in particular its harder 3-momentum dependence, found here could have significant ramifications for the phenomenology of open HF probes in URHICs. In a recent work~\cite{He:2019vgs} a good description of $D$, $D_s$ and $\Lambda_c$ observables in heavy-ion collisions has been achieved using the $T$-matrix based transport coefficients from Refs.~\cite{Riek:2010fk,Huggins:2012dj}, which are based on the internal-energy ($U$) as a potential proxy but with an extra $K$ factor of about 1.6. The pertinent results for $A(p,T)$ (with $K=1.6$) are slightly larger than the ones from the SCS with $\chi=1$ at low momentum, but much larger at higher momenta. However, with our new $\chi=0.6$ results, the low-momentum deficit can be overcome, while they still fall below the high-momentum results of the $U$-potential with $K=1.6$. Yet, the inclusion of radiative processes, as computed within the $T$-matrix approach in Ref.~\cite{Liu:2020dlt} could result in a total transport coefficients that are quite comparable to the one employed in Ref.~\cite{He:2019vgs}, without the need of any phenomenological adjustments.

\section{Conclusions}
\label{sec_concl}
We have augmented the thermodynamic $T$-matrix approach to include the effects of spin-dependent interactions between heavy quarks, including spin-orbital, spin-spin and tensor contributions, as part of the more general objective to assess 1/$M_Q$ corrections. Toward this end we have utilized the Breit-Fermi Hamiltonian to derive these interactions in the context of the Cornell potential as the two-body interaction kernel for the $T$-matrix equation. When benchmarking these interactions using the experimentally observed splittings in vacuum quarkonium spectroscopy, 
we have found that, in accordance with previous studies, a moderate admixture of a Lorentz-vector component in the confining potential allows for a much improved description especially in the charmonium sector. We have then implemented the amended interaction kernel into our quantum many-body approach for the QGP. While the spin-dependent interactions themselves are expected to be of minor importance (and therefore have been neglected), the vector component of the confining potential turns out to be rather significant. After selfconsistently refitting the in-medium HQ free energies and the QGP EoS under the inclusion of the vector component, quantitative modifications of the QGP properties toward shorter distances (larger momenta) were found. A strong broadening of the thermal-parton spectral functions persists to higher 3-momenta as a consequence of an increased interaction strength in the thermodynamic 2-body scattering amplitudes at larger momenta. The harder amplitudes and spectral functions are a consequence of the relativistic corrections induced by the vector part of the confining interaction, as opposed to a purely scalar interaction. This suggests that the nature of the confining force in the QCD vacuum has an impact on the properties of the strongly-coupled QGP, with liquid properties that extend to higher resolution scales compared to a purely scalar confining force. Finally, we have applied the modified set-up to calculate the friction coefficient of charm quarks. As compared to the results with a purely scalar string potential, a slightly larger relaxation rate is found at small momentum (and a pertinent decrease in the diffusion coefficient), but a much larger increase of a factor of $\sim$2-3 (or more) at momenta of around 5\,GeV (and above). These are promising features to make a significant step forward in achieving a quantitative description of HF diffusion in heavy-ion collisions at RHIC and the LHC based on microscopically and non-perturbatively calculated transport coefficients. Work in this direction is in progress.


\acknowledgments
This work has been supported by the U.S. National Science Foundation under grant nos. PHY-1913286 and PHY-2209335, and by the U.S. Department of Energy, Office of Science, Office of Nuclear Physics through the Topical Collaboration in Nuclear Theory on \textit{Heavy-Flavor Theory (HEFTY) for QCD Matter} under award no.~DE-SC0023547.

\bibliography{refnew}

\end{document}